\documentclass[12pt, draftclsnofoot, onecolumn]{IEEEtran}
\usepackage{cite}
\usepackage{graphicx}
\usepackage{amsmath}
\usepackage{array}
\usepackage{mdwmath}
\usepackage{amssymb}
\usepackage{mdwtab}
\usepackage{subfigure}
\usepackage{stfloats}
\usepackage{amsthm}
\usepackage{threeparttable}
\usepackage{color}
\usepackage{ulem}
\usepackage{epstopdf}

\makeatletter
\newif\if@restonecol
\makeatother

\usepackage[linesnumbered,ruled,vlined]{algorithm2e}
\usepackage{algpseudocode}
\usepackage{amsmath}
\usepackage{contour}
\usepackage{ulem}

\begin{document}
\title{ {\fontsize{20}{25}\selectfont On the IRS Deployment in Smart Factories Considering Blockage Effects: Collocated or Distributed?}
}
\author{Yixin Zhang, 
	Saeed R. Khosravirad, 
    Xiaoli Chu, 
	Mikko A. Uusitalo
		
\thanks{Yixin Zhang (yixin.zhang@bupt.edu.cn) is  with  the Key Laboratory of Universal Wireless Communications, Ministry of Education, Beijing University of Posts and Telecommunications, China, and with the Department of Electronic and Electrical Engineering, University of Sheffield, UK.}
\thanks{Saeed R. Khosravirad (saeed.khosravirad@nokia-bell-labs.com) is with the Nokia Bell Labs, Murray Hill, NJ, USA.}
\thanks{Xiaoli Chu (x.chu@sheffield.ac.uk) is with the Department of Electronic and Electrical Engineering, University of Sheffield, UK.}
\thanks{Mikko A. Uusitalo (mikko.uusitalo@nokia-bell-labs.com) is with the Nokia Bell Labs, Espoo, Finland.}
\thanks{This work is supported in part by the Fundamental Research Funds for the Central Universities under Grant 2023RC18, and in part by the European Union’s Horizon 2020 Research and Innovation Programme under Grant 766231.}
}
\markboth{Arxiv}%
{Shell \MakeLowercase{\textit{et al.}}: Bare Demo of IEEEtran.cls for Journals}
\maketitle
\vspace{-10pt}
\begin{abstract}
In this article, we study the collocated and distributed deployment of intelligent reflecting surfaces (IRS) for a fixed total number of IRS elements to support enhanced mobile broadband (eMBB) and ultra-reliable low-latency communication (URLLC) services inside a factory. We build a channel model that incorporates the line-of-sight (LOS) probability and power loss of each transmission path, and propose three metrics, namely, the expected received signal-to-noise ratio (SNR), expected finite-blocklength (FB) capacity, and expected outage probability, where the expectation is taken over the probability distributions of interior blockages and channel fading.
The expected received SNR and expected FB capacity for extremely high blockage densities are derived in closed-form as functions of the amount and height of IRSs and the density, size, and penetration loss of blockages, which are verified by Monte Carlo simulations. Results show that deploying IRSs vertically higher leads to higher expected received SNR and expected FB capacity. By analysing the average/minimum/maximum of the three metrics versus the number of IRSs, we find that for high blockage densities, both eMBB and URLLC services benefit from distributed deployment; and for low blockage densities, URLLC services benefit from distributed deployment while eMBB services see limited difference between collocated and distributed deployment.

\end{abstract}

\section{Introduction}
Industrial 4.0 is envisioned to automate or upgrade various industry processes \cite{Indus,Indus2}, building on enhanced wireless connectivity offered by the fifth-generation (5G) and beyond 5G (B5G) wireless communications technology \cite{5GACIA,6G}.
The 3rd Generation Partnership Project (3GPP) has defined use cases for the ``Factory of the Future", where the wireless performance requirements are mainly specified for two of the 5G service types \cite{5GACIA, next}: enhanced mobile broadband (eMBB) services, such as high definition videos, virtual reality, and augmented reality, which require high data rates and broad bandwidths; and ultra-reliable low-latency communications (URLLC) services, e.g., for industrial automation and autonomous driving, which require an end-to-end latency of 0.5-500 ms and a communication service availability ranging from 99.9\% to 99.999999\%. 

Future smart factories will comprise industry equipment and furniture, as well as numerous sensing, computing, and actuating devices, which may become blockages to wireless signal propagation. Multi-hop communication \cite{diversity,stogeo}, distributed multiple-input multiple-output (D-MIMO) \cite{DMIMO1}, and multi-point coordination \cite{linklevel} schemes have been proposed to facilitate communications in the presence of blockages and clutter. 
Recently, intelligent reflecting surfaces (IRSs), which are capable of manipulating wireless propagation environments, have emerged as a promising technology of realising cost-effective and energy-efficient wireless communications in smart factories \cite{pathplanning, symbiotic, SRE}. However, how IRSs should be deployed (e.g., collocated or distributed) in a smart factory to best support diverse eMBB and URLLC services has not been sufficiently studied.

\subsection{Related Work}

\textit{Existing works on millimetre wave (mmWave) propagation in factories:} 
The channel characteristics of mmWave signal propagation have been widely studied through measurement campaigns. The authors in \cite{comparative} provided the results of the path loss and Rician K-factor centred at 28 GHz with a bandwidth of 800 MHz in indoor factory environments. The authors in \cite{pathdirect} presented the results of the path gain and the azimuthal angular power distribution at 28 GHz with a receive bandwidth of 20 kHz in a factory. The authors in \cite{characterisation} analysed the path loss and line-of-sight (LOS) probability in both light and heavy industry factories under the candidate frequencies of 28 and 60 GHz for licensed- and unlicensed-band communication, respectively, and suggested that mmWave communications in indoor industrial scenarios should take into account the specific geometry of the indoor environment.
The authors in \cite{iiotsurvey} noted that the bands 24.25-27.5 GHz and the bands above 52.6 GHz are the candidate bands for Internet-of-things (IoT) applications with high priority, and claimed that the precise models of path loss, LOS probability, root mean-square (RMS) delay spread, and angular spread should consider the antenna height, clutter density, factory volume, and the total surface.



\textit{Existing works on distributed cooperative IRSs:}
In order to overcome the rank deficiency of IRS-assisted wireless networks, multiple IRSs are required to offer spatial multiplexing gains.
The authors in \cite{covpro} found that deploying a large number of finite-size IRSs achieves a higher coverage probability than deploying a small number of large IRSs for a single-input single-output (SISO) system under correlated Rayleigh fading channels.
The authors in \cite{analysis} derived the outage probability of a multi-IRS-assisted SISO system under Rician fading and showed that when the LOS components are stronger than the non-LOS (NLOS) ones, the minimal outage probability, attained through the phase shift alignment between the direct and reflected links, decreases with the number of IRSs and/or the number of elements per IRS.
The authors in \cite{perana} derived the asymptotic outage probability and average symbol error rate for a distributed IRS aided SISO system under Nakagami-$m$ fading, assuming that the direct and reflected signals are constructively added at the receiver, and unveiled that the achievable diversity order linearly increases with the number of distributed IRSs and the number of elements per IRS.
The authors in \cite{statisticalCSI} analysed the ergodic rate of a multiple-input single-output (MISO) system assisted by multiple distributed IRSs considering the impact of channel estimation errors and revealed that the distributed deployment of IRSs is superior to the centralised deployment due to the increased LOS probability of the paths.

\textit{Existing works on large-scale IRS deployment:} 
In the large-scale IRS deployment \cite{spathr,spadis,randomly,cost}, only one of the multiple randomly deployed IRSs is selected/associated to serve the user equipment (UE).
The authors in \cite{spathr} characterised the spatial throughput of a single-cell multiuser system, and revealed that the system spatial throughput increases when fewer IRSs each with more reflecting elements are deployed, but at the cost of degraded user-rate fairness.
The authors in \cite{spadis} compared the SISO system aided by multiple distributed IRSs or relays, where a user is served by the relay or IRS that leads to the highest received signal-to-noise ratio (SNR), and concluded that the IRSs outperform relays in terms of outage probability and energy efficiency especially when each IRS is equipped with a large number of IRS elements or the IRSs are more densely deployed.
The authors in \cite{randomly} studied an outdoor cellular network, where the base stations (BSs), IRSs, and blockages are all randomly distributed on a 2D plane, UEs are associated with the BS that offers the lowest average path loss, and a fraction of the blockages are equipped with IRSs.
Their results show that a large-scale deployment of IRSs can reduce the blind-spot areas and that coating the blockages at chosen locations with IRSs can reduce the required density of IRSs.
The authors in \cite{cost} studied the coverage probability of an IRS-assisted LOS mmWave network, and concluded that deploying more small IRSs outperforms deploying fewer large IRSs for small-cell networks with a high user density, e.g., malls and airports.



In smart factories, IRSs are typically deployed on surrounding walls \cite{wallIRS,wallIRS2}.
We note that the existing works on IRS deployment have not sufficiently studied the effects of interior blockages on wireless signal propagation in 3D indoor factory environments. For instance, the widely adopted channel models either assume that the direct link between the BS and the UE is completely blocked (i.e., the blockages are impenetrable \cite{blockage}) while neglecting the possible residual signal strength passing through the blockages \cite{spadis,randomly,cost} or assume that both the BS-IRS and IRS-UE links are LOS \cite{analysis,perana,statisticalCSI,spathr,spadis}, which however may not always be the case.

\subsection{Contributions}

%



To fill the aforementioned gaps, in this article, we investigate the collocated and distributed IRS deployment strategies in smart factories for a fixed total number of passive reflecting elements, taking into account the effects of interior blockages and different quality-of-service (QoS) requirements of eMBB and URLLC services. The main contributions are summarised as follows:
\begin{itemize}
	\item We develop a 3D system model for a cubic factory, where a BS is deployed at the centre of the ceiling, a typical UE randomly locates in the blind spot area behind a big tall shelf in the middle of the factory, multiple IRSs are deployed on the walls surrounding the blind spot area, and there are interior blockages with random locations and heights modelled following a Poisson point process (PPP) and a uniform distribution, respectively. For the considered typical UE of an arbitrary location in the blind spot area, we derive the LOS probability for each IRS-UE link. 
	\item Assuming that each IRS-UE link is either LOS or NLOS independently, we define every distinct combination of LOS/NLOS status of all the IRS-UE links as a blockage case and obtain the probability of occurrence for each blockage case. For an arbitrary blockage case, we derive the expressions of the direct channel from the BS to the typical UE and the indirect channel from the BS via the multiple IRSs to the typical UE.
	\item To facilitate the performance comparison between collocated and distributed IRS deployment strategies, we design three UE location-specific performance metrics, i.e., the expected received SNR, the expected finite-blocklength (FB) capacity, and the expected outage probability, where the expectation is taken over all possible blockage cases and channel fading. The first two metrics are relevant to eMBB services that require a high received SNR and a high data rate. The last two metrics are relevant to URLLC services that require a high FB capacity and a low outage probability. 
	\item The expected received SNR and the expected FB capacity for extremely high blockage densities are derived in closed forms as functions of the number and height of IRSs, the density and size of blockages, and the penetration power loss per blockage. The analytical expressions are verified by Monte Carlo simulations. Our analytical and numerical results show that deploying IRSs higher on the walls results in a higher expected received SNR and a higher expected FB capacity.
	\item Based on the analytical and simulation results, we provide following insights into the IRS deployment in a smart factory: For URLLC services,
	distributed IRSs outperform collocated IRSs in terms of the minimum expected FB capacity and the maximum expected outage probability across all possible UE locations for both high and low blockage densities. 
	For eMBB services, collocated or distributed IRSs achieve similar values of UE-location averaged expected received SNR and similar values of UE-location averaged expected FB capacity for low blockage densities, but distributed IRSs offer a higher UE-location averaged expected received SNR and a higher UE-location averaged expected FB capacity than collocated IRSs for high blockage densities. 
\end{itemize}

The rest of the article is organised as follows.
Section II introduces the system model for IRS-assisted downlink
communications in a factory. In Section III, three performance metrics for eMBB and URLLC services are defined and derived. In Section IV, the numerical and  simulation results are presented. Finally, Section V concludes this article with guidelines on IRS deployment in smart factories.

\section{System Model}

\subsection{Factory Environment}

\begin{figure} [!t]
	\centering
	\includegraphics[width=4in]{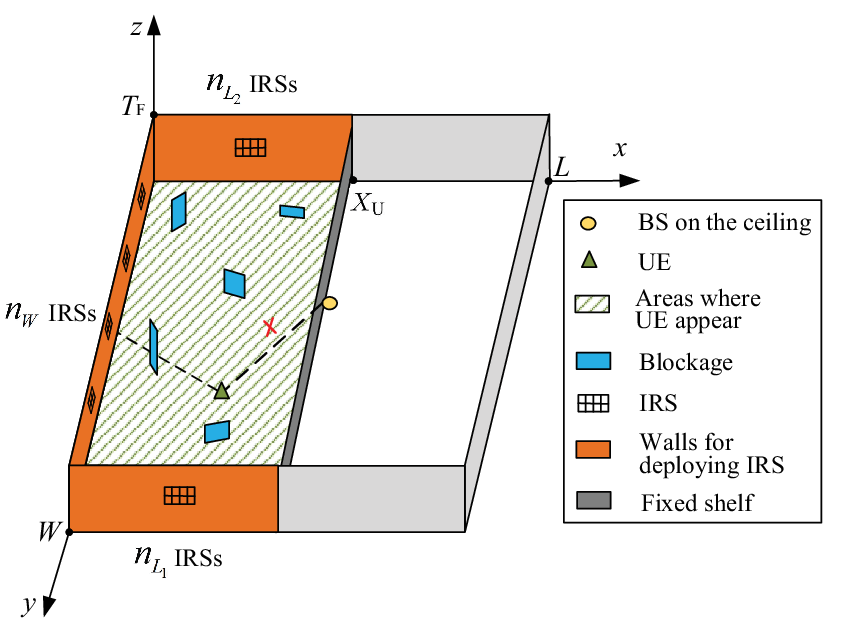}
	\caption{A cubic $L \times W\times T_{\text{F}}$ m$^3$ factory, where the BS is at the centre of the ceiling, blockages are randomly oriented and distributed on the ground, the UE behind the shelf is served with the aid of the IRS(s) on the 3 side walls.}
\end{figure}

As shown in Fig. 1, we build a 3D Cartesian coordinate system inside a factory of size $L \times W\times T_{\text{F}}$ m$^3$, where $L$, $W$ and $T_{\text{F}}$ are the length, width, and ceiling height of the factory, respectively. A BS is located at the centre of the ceiling with the coordinate of $[L/2,W/2,T_{\text{F}}]$. A fixed blocker, e.g., a big tall shelf, whose length and height are close to $W$ and $T_{\text{F}}$ respectively, is located in the 2D plane $x=X_{\text{U}}$ parallel to the wall of size $W \times T_{\text{F}}$ and near the BS, i.e., $X_{\text{U}}$ is slightly less than $L/2$, thus blocking and causing a power loss of $\omega$ to the link between the BS and any UE located behind the shelf. Without loss of generality, we consider a UE located in the above described blind spot of the BS's coverage area, and denote the UE's coordinate by $[x_{\text{U}},y_{\text{U}},T_{\text{U}}]$, where $x_{\text{U}} \in (0, X_{\text{U}})$, $y_{\text{U}} \in (0, W)$, and $T_{\text{U}}$ denotes the height of the UE's antenna. For analytical tractability, both the BS and UE are assumed to be equipped with a single antenna.

In order to enhance the wireless communication quality in the blind spot behind the shelf, $M$ cooperative IRSs are deployed on the three walls surrounding the blind spot, i.e., one wall of size $W \times T_{\text{F}}$ on the opposite side of the shelf with respect to the BS and two walls each of size $L/2 \times T_{\text{F}}$ perpendicular to the shelf. The $M$ IRSs are deployed at the same height of $h$ and evenly share a total of $N$ reflective elements, i.e., each IRS has $\frac{N}{M}$ elements that are arranged into an $N_{\text{v}} \times N_{\text{h}}$ planar array with the inter-element spacing of $l$ m. For simplicity, we assume that $\frac{N}{M}$ is an integer.

Let ${n_{{L_1}}}$ and ${n_{{L_2}}}$ denote the number of IRSs deployed on the two walls perpendicular to the shelf, and ${n_W}$ denote the number of IRSs deployed on the wall parallel to the shelf, subject to ${n_{{L_1}}} + {n_{{L_2}}} + {n_W} = M$. Letting $\tau=\frac{W}{X_{\text{U}}} $ denote the aspect ratio of the blind spot area, the number of IRSs evenly deployed on each of the above mentioned three walls is given as follows.

If $\tau\geq 1$, then
\begin{equation}
\begin{aligned}
{n_{{L_1}}} &= {n_{{L_2}}} = \left\lfloor {\frac{M}{{\tau + 2}}} \right\rfloor,\\ {n_W} &= M - 2{n_{{L_1}}}.
\end{aligned}
\end{equation}

If $\tau<1$, then 
\begin{equation}
\begin{aligned}
{n_W} &= \left\lfloor {\frac{{\tau M}}{{2 + \tau}}} \right\rfloor,\\
	{n_{{L_1}}} &= {n_{{L_2}}} = \frac{{M - {n_W}}}{2}, {\text{if \ }} M - {n_W} {\text{ \ is even}},\\
	{n_{{L_1}}} &= \frac{{M \!-\! {n_W} \!+\! 1}}{2},{n_{{L_2}}} = \frac{{M \!-\! {n_W} \!-\! 1}}{2}, {\text{if \ }} M - {n_W} {\text{ \ is odd,}}
\end{aligned}
\end{equation}
where $\left\lfloor \cdot \right\rfloor$ denotes the floor function.

Accordingly, the deployment location of the $m$th IRS ${\bf{q}}_m$ is given by
\begin{equation}
\begin{aligned}
{{\bf{q}}_m} = \left\{\begin{array}{l}
\left( {0,\frac{m}{{{n_W} + 1}}W,h} \right), m=1, 2, ..., n_W, \\
 \left( {\frac{{m-n_W}}{2\left({{n_{{L_1}}} + 1}\right)}L,W,h} \right), m={n_W}+1,  ..., {n_W}+{n_{{L_1}}},\\ 
\left( {\frac{m-n_W-{n_{{L_1}}}}{2\left({{n_{{L_2}}} + 1}\right)}L,0,h} \right), m={n_W}+{n_{{L_1}}}+1, ..., M.	
\end{array} \right.
\end{aligned}
\end{equation}

Given the coordinates of the BS, the UE, and the $m$th IRS for $m=1,2,..,M$, we can calculate the distance from the BS to the UE, the distance from the $m$th IRS to the UE, and the distance from the BS to the $m$th IRS, which are denoted by $d_0$, $d_m$, and 
$D_m$, respectively.

\subsection{UE-specific Channel Model Considering Blockage Effects}
We adopt the stochastic geometry based random blockage model \cite{blockage}, where rectangular screen blockages \cite{3gpp}, whose locations are defined by the centres of their bottom lines, are distributed following a PPP with the density of $\lambda_{\text{B}}$ (blockage/m$^2$) and with their orientations uniformly distributed in $[0, \pi]$.
The height of each blockage follows an independent uniform distribution in the range $[T_{\text{U}}, T_{\text{B}}]$, where $T_{\text{B}}$ denotes the maximum height of the blockages. We assume $T_{\text{B}} \le h \le T_{\text{F}}$ so that the links between the BS and the IRSs are always LOS. 
The blockages have an identical width of $R_{\text{B}}$ m.  

The direct channel from the BS to the considered UE is given by
	\begin{equation}\label{channeldirect}
		{{f}_0} = \sqrt {{\beta _0}\omega {\upsilon ^{{B_{0}}}}} {{{f}}_{\text{bu}}},
	\end{equation}
	where
	\begin{equation}\label{PL0}
		\beta_0=\frac{{{G_{\text{T}}}{G_{\text{R}}}{\mu ^2}}}{{{{\left( {4\pi {d_0}} \right)}^2}}},
	\end{equation}
	denotes the free-space path loss along the the BS-UE link, $v^{B_{0}}$ denotes the total loss caused by the random blockages along the BS-UE link, in which $v$ denotes the attenuation of the signal strength caused by one blockage, and $B_0$ denotes the number of blockages intersecting the BS-UE link.
	Since the direct link from the BS to the UE is always blocked due to the fixed shelf, the channel between them sees Raleigh fading, i.e., ${{f}}_{{\rm{bu}}}\sim \mathcal{CN}\left( 0, 1 \right)$.

For analytical tractability, we assume that the blockages affect each IRS-UE link independently. This assumption holds for most UE locations where the IRS-UE links are independent to one another \cite{blockage}. 
Thus, each of the $M$ IRS-UE links can be either LOS or NLOS independently, leading to $2^M$ different combinations of LOS/NLOS status of the $M$ IRS-UE links. Each of the $2^M$ combinations is referred to as a blockage case.

A specific blockage case can be represented by a set $\Xi$, which contains the indices of the LOS IRS-UE links in the corresponding blockage case, where  $\Xi  \subset \Lambda$, $\Lambda  = \left\{ 1,2,...,M \right\}$, and $0 \le \left| \Xi  \right| \le M$. Accordingly, the set $\Lambda \backslash \Xi$  contains the indices of the NLOS IRS-UE links in the blockage case represented by $\Xi$. Thus, the blockage case $\Xi$ occurs with the probability
\begin{equation}\label{pro}
{\zeta _\Xi } = \prod\limits_{m \in \Xi }^{} {{p_m}} \prod\limits_{w \in \Lambda \backslash \Xi }^{} {(1 - {p_w})}.
\end{equation}

Given the random spatial distribution of blockages, the number of blockages intersecting the $m$th IRS-UE link is a Poisson random variable \cite{blockage}, denoted by $B_m$. For a specific blockage case $\Xi$, the number of blockages intersecting the $m$th IRS-UE link is denoted by $B_{m,\Xi}$, which is a realisation of $B_m$. Note that $B_{m, \Xi}=0$ only when $m \in \Xi$, while $B_{m, \Xi}$ is a positive integer when $m \in \Lambda \backslash \Xi $. 

\textbf{Lemma 1}: Letting $d_{\text{2D},0}$ and $d_{\text{2D},m}$ denote the horizontal distances in the XOY plane of the BS-UE link and the $m$th IRS-UE link, respectively, the expected number of blockages intersecting the BS-UE link and the $m$th IRS-UE link are, respectively, given by
	
	\begin{equation}\label{blocker0}
		E(B_0)= \frac{{\left( T_{\text{B}} - {T_{\text{U}}} \right)}}{\left( T_{\text{F}} - T_{\text{U}} \right)} \frac{{\lambda _{\text{B}}}{R_{\text{B}}}{d_{\text{2D},0}}}{\pi },
	\end{equation}
	\begin{equation}\label{blockerm}
		E(B_m)=\frac{{\left( T_{\text{B}} - {T_{\text{U}}} \right)}}{\left( h - T_{\text{U}} \right)} \frac{{\lambda _{\text{B}}}{R_{\text{B}}}{d_{{\rm{2D}},m}}}{\pi }, m=1,2,..,M,
	\end{equation}
	where $E(\cdot)$ denotes the expectation; and the LOS probability of the $m$th IRS-UE link is given by
	\begin{equation}\label{losm}
		p_m=\exp({-E(B_m)}), m=1,2,..,M.
	\end{equation}
	
	\begin{IEEEproof}
		See Appendix A.
	\end{IEEEproof}

For the blockage case $\Xi$, the indirect channel from the BS via the $M$ IRSs to the considered UE is given by
\begin{equation}\label{channelIRS}
{{f}_{\Xi}} = \sum\limits_{m = 1}^M {\sqrt {{\beta _m}{\upsilon ^{{B_{m,\Xi}}}}}} {{\bf{f}}_{{\text{ru}},m}}{\Theta _m}{{\bf{F}}_{{\text{br}},m}},
\end{equation}
where
\begin{equation}\label{PLm}
\beta_m= \frac{{{G_{\text{T}}}{G_{\text{R}}}{\mu ^2}}}{{{{\left( {4\pi } \right)}^3}}}{\left( {\frac{l}{{{D_m}{d_m}}}} \right)^2}{\cos ^2}\left( {{\varphi _m}} \right), \forall m=1,2,...,M,
\end{equation}
denotes the free-space path loss along the $m$th IRS-UE link \cite{PL1}, in which $\mu$ denotes the wavelength, ${G_{\text{T}}}$ and ${G_{\text{R}}}$ denote the BS transmit and UE receive antenna gain in dBi, respectively,  and $\varphi_m$ denotes the incident angle with respect to the $m$th IRS; $v^{B_{m,\Xi}}$ denotes the total loss caused by the blockages along the $m$th IRS-UE link under blockage case $\Xi$; $\Theta_m \in {\mathbb{C}^{N/M \times N/M}}$ denotes the phase shift matrix of the $m$th IRS and is given by
\begin{equation}\label{phase}
{\Theta_m}  = {\rm{diag}}[{e^{j{\theta _{m,1}}}},{e^{j{\theta _{m,2}}}},...,{e^{j{\theta _{m,N/M}}}}],
\end{equation}
where $\theta _{m,n}$ denotes the phase shift at the $n$th element of the $m$th IRS, $m=1,2,...,M$, $n=1,2,...,N/M$, and ${\rm{diag}}[\cdot]$ denotes the diagonal matrix.

The LOS channel between the BS and the $m$th IRS is given by \cite{jointactpass}
\begin{equation}
{{\bf{F}}_{{\rm{br}},m}} \in {\mathbb{C}^{{N_{\rm{h}}}{N_{\rm{v}}} \times 1}} = {{\bf{w}}^H}\left( {\delta _{{\rm{br}},m}^{(h)},\delta _{{\rm{br}},m}^{(v)},{N_\text{h}},{N_\text{v}}} \right),
\end{equation}
where ${\bf{w}}\left( {{x^{(h)}},{x^{(v)}},A,B} \right)$ is the row vectorisation of ${\frac{1}{{AB}}{\bf{W}}\left( {{x^{(h)}},{x^{(v)}},A,B} \right)}$,
the $(a, b)$th element of ${{\bf{W}}\left( {{x^{(h)}},{x^{(v)}},A,B}\right)}$ is $e^{j\varpi \left( {{x^{(h)}},{x^{(v)}},a,b} \right)}$ for $a \!=\! 1,2,...A$, $b = 1,2,...B$,
and
\begin{equation}
\begin{array}{l}
\varpi \left( {{x^{(h)}},{x^{(v)}},a,b} \right)
= \frac{{2\pi l\sin {x^{(v)}}\left( {\left( {a \!-\! 1} \right)\cos {x^{(h)}} + \left( {b \!-\! 1} \right)\sin {x^{(h)}}} \right)}}{\lambda },
\end{array}
\end{equation}
where $\delta _{{\rm{br}},m}^{(h)}$ and $\delta _{{\rm{br}},m}^{(v)}$ denote the horizontal and vertical angle of arrival (AOA) from the BS to the $m$th IRS, respectively, and $(\cdot)^H$ denotes the conjugate transpose.

The channel between the $m$th IRS and the UE, ${\bf{f}}_{{\rm{ru}},m} \in {\mathbb{C}^{1 \times N/M}}$, is assumed to follow Rician fading, i.e.,
\begin{equation}
\begin{aligned}
{{\bf{f}}_{{\rm{ru}},m}}  &= \sqrt {\frac{{{K_{{\rm{ru}},m}}}}{{1 + {K_{{\rm{ru}},m}}}}} {{\bf{\bar f}}_{{\rm{ru}},m}} + \sqrt {\frac{1}{{1 + {K_{{\rm{ru}},m}}}}} {{\bf{\tilde f}}_{{\rm{ru}},m}},
\end{aligned}
\end{equation}
where ${K_{{\rm{ru}},m}}$ denotes the Rician factor and is given by \cite{comparative}
\begin{equation}
{K_{{\rm{ru}},m}} = \left\{ {\begin{array}{*{20}{l}}
		{7.34 - 0.046{d_m} \ ({\rm{dB}}), {\rm{if \ }} m \in \Xi,}\\
		{0, {\rm{if \ }} m \in \Lambda \backslash \Xi,}
\end{array}} \right.
\end{equation}
${{\bf{\bar f}}_{{\rm{ru}},m}}$ and ${{\bf{\tilde f}}_{{\rm{ru}},m}}$ denote the LOS and NLOS components, respectively, 
\begin{equation}
{{\bf{\bar f}}_{{\rm{ru}},m}} = {\bf{w}}\left( {\delta _{{\rm{ru}},m}^{(h)},\delta _{{\rm{ru}},m}^{(v)},{N_{\rm{h}}},{N_{\rm{v}}}} \right),
\end{equation}
in which $\delta _{{\rm{ru}},m}^{(h)}$ and $\delta _{{\rm{ru}},m}^{(v)}$ denote the horizontal and vertical angle of departure (AOD) from the $m$th IRS to the UE, respectively, each element of $ {{{{\bf{\tilde f}}}_{{\rm{ru}},m}}}$ is assumed to be i.i.d. following Gaussian distribution with a zero mean and unit variance. 

In order to enable constructive received signal combining at the UE, the optimal phase shift matrices of the $M$ IRSs given in \eqref{phase} are configured based on ${{f}}_{{\rm{bu}}}$, ${{\bf{f}}_{{\rm{ru}},m}}$, and ${{{\bf{F}}_{{\rm{br}},m}}}$ \cite{perana} as follows:
\begin{equation}\label{IRSphase}
\begin{aligned}
\theta _{m,n} &= \arg \left( { {{{{f}}_{{\rm{bu}}}}} } \right) - \arg \left( f_{{\rm{ru}},m,n}  \right) - \arg \left(  F_{{\rm{br}},m,n} \right),\\ & m=1,2,...,M, n=1,2,...,N/M,
\end{aligned}
\end{equation}
where $f_{{\rm{ru}},m,n} $ and $F_{{\rm{br}},m,n}$ denote the $n$th element of ${{\bf{f}}_{{\rm{ru}},m}}$ and ${{\bf{F}}_{{\rm{br}},m}}$, respectively.

Accordingly, the signal received at the UE under blockage case $\Xi$ is given by
	\begin{equation}
	r =  {{f}_0}t+{{f}_{\Xi}}t + n,
	\end{equation}
where $r$, $t$, $n$ denote the received signal, the transmitted signal with the transmission power of $P_{\text{T}}$ (dBm), and the additive white Gaussian noise with the power of ${P_{\text{W}}}$, respectively, ${{f}_0}$ and ${{f}_{\Xi}}$ are given in \eqref{channeldirect} and \eqref{channelIRS}, respectively. Given the noise floor of $\sigma$ (dB) and the bandwidth of $Z$ (MHz), ${P_{\text{W}}}$ is given by ${P_{\text{W}}}=-174+\sigma+10 \log_{10} Z$ (dBm). Then, the transmit SNR is given by $\rho=P_{\text{T}}-P_{\text{W}}$ (dB).

Hence, the received SNR under blockage case $\Xi$ is given by
\begin{equation}
\begin{aligned}
{{\gamma _\Xi }} &\!=\!\rho{{\left| {{f}_0}+{{f}_{\Xi}} \right|}^2}\\ &\!=\! \rho{\left( {\sqrt {{\beta _0}\omega {\upsilon ^{{B_0}}}} \left| {{f_{{\rm{bu}}}}} \right| \!+\! \sum\limits_{m = 1}^M {\sqrt {{\beta _m}{\upsilon ^{{B_{m,\Xi}}}}} \sum\limits_{n = 1}^{N/M} {\left| f_{{\rm{ru}},m,n} \right|} } } \right)^2}.
\end{aligned}
\end{equation}


For URLLC services in smart factories, typical packets are very short in order to meet the stringent latency requirement \cite{5GACIA}, and the channel capacity is evaluated using the FB channel capacity. Under blockage case $\Xi$, the FB channel capacity in bit/s/Hz is given by \cite{wallIRS}
\begin{equation}
{C_{ \Xi  }^{\text{FB}}} = {\log _2}\left( {1 + \gamma _\Xi} \right) - \sqrt {\frac{1}{S} - \frac{1}{{S{{\left( {1 + \gamma _\Xi } \right)}^2}}}} \frac{{{Q^{ - 1}}\left( \varepsilon  \right)}}{{\ln 2}},
\end{equation}
where $S$ denotes the blocklength in nat, $\varepsilon$ denotes the decoding error probability, and $Q^{-1}$ denotes the inverse Q function.

The outage probability under blockage case $\Xi$ is given by
\begin{equation}
\begin{aligned}
P_{ \Xi  } &= \Pr \left( {{{\log }_2}\left( {1 + \rho{{\left|  {{f}_0}+{{f}_{\Xi}} \right|}^2}} \right) < R} \right) \\ &= \Pr \left( {{\left|  {{f}_0}+{{f}_{\Xi}} \right|}^2} < \frac{{2^{R} - 1}}{\rho} \right) ,
\end{aligned}
\end{equation}
where $R$ denotes the capacity threshold in bit/s/Hz.

\section{Performance Metrics}
In this section, we define three new UE location-specific performance metrics, i.e., the expected received SNR, the expected FB capacity, and the expected outage probability, which can be used to compare the collocated and distributed IRS deployment schemes in smart factories. The first two metrics are relevant to eMBB services, while the last two metrics are relevant to URLLC services. For extremely high blockage densities, the expected received SNR and the expected FB capacity are derived in closed forms.

\subsection{Expected Received SNR}
Based on the probability $\zeta_\Xi$ of blockage case $\Xi$ in (\ref{pro}) and the received SNR under $\Xi$ in (20), the expected received SNR for all possible $\Xi$ at a specific UE location is defined as
\begin{equation}\label{SNRdef}
\gamma  = \sum\limits_\Xi ^{} {{\zeta _\Xi }E\left( {{\gamma _\Xi }} \right)}.
\end{equation}

\textbf{Theorem 1}: In extremely high blockage density scenarios (i.e., $\lambda_b$ is large but finite), 
$\gamma$ approaches $E\left( {{\gamma _\varnothing }} \right)$, where $\gamma_\varnothing = \gamma_\Xi$ for $\Xi$ being a null set corresponding to the high-density blockage case that has no LOS IRS-UE link at all, and $E\left( {\gamma _\varnothing} \right)$ is given as a function of $M$, $h$, $\lambda_{\rm{B}}$, $R_{\rm{B}}$, and $v$ in (24).

	\begin{equation}\label{EMh}
	\begin{aligned}
	E\left( {{\gamma _\varnothing}} \right)  &\!=\! \frac{{\rho {G_{\text{T}}}{G_{\text{R}}}{\mu ^2}}}{{16{\pi ^2}}} \times\\
	&\left( \begin{array}{l}
	\frac{\omega }{{d_0^2}}\exp \left( {  \frac{{-\left( {{T_{\rm{B}}} - {T_{\rm{U}}}} \right){\lambda _{\text{B}}}{R_{\rm{B}}}{d_{{\rm{2D}},0}}\left( {1 - v} \right)}}{{\left( {{T_{\rm{F}}} - {T_{\rm{U}}}} \right)\pi }}} \right)\\
	+ \frac{{Nl\sqrt {\pi \omega } }}{{4M{d_0}}}\exp \left( { \frac{{- \left( {{T_{\rm{B}}} - {T_{\rm{U}}}} \right){\lambda _{\text{B}}}{R_{\text{B}}}{d_{{\rm{2D}},0}}\left( {1 - \sqrt v } \right)}}{{\left( {{T_{\rm{F}}} - {T_{\rm{U}}}} \right)\pi }}} \right)\sum\limits_{m = 1}^M {\frac{{\cos {\varphi _m}}}{{{D_m}{d_m}}}\exp \left( { \frac{{-\left( T_{\text{B}} - {T_{\text{U}}} \right){{\lambda_{\rm{B}}}{R_{\text{B}}}{d_{{\rm{2D}},m}}\left( {1 - \sqrt{v}} \right)}}}{\left( h - T_{\text{U}} \right)\pi} } \right)} \\
	+ \frac{{{N^2}l^2}}{{16M^2}}\sum\limits_{m = 1}^M {\sum\limits_{p = 1,p \ne m}^M {\frac{{\cos {\varphi _m}}}{{{D_m}{d_m}}}\frac{{\cos {\varphi _p}}}{{{D_p}{d_p}}}\exp \left( {  \frac{{{-\left( T_{\text{B}} - {T_{\text{U}}} \right)\lambda _{\text{B}}}{R_{\text{B}}}\left( {{d_{{\rm{2D}},m}} + {d_{{\rm{2D}},p}}} \right)\left( {1 - \sqrt v } \right)}}{\left( h - T_{\text{U}} \right)\pi }} \right)} } \\
	+ \frac{{Nl^2}}{{4\pi M }}\left( {1 - \frac{\pi }{4}}+ \frac{N }{M} \right)\sum\limits_{m = 1}^M {{{\left( {\frac{{\cos {\varphi _m}}}{{{D_m}{d_m}}}} \right)}^2}\exp \left( {  \frac{{{-\left( T_{\text{B}} - {T_{\text{U}}} \right)\lambda _{\text{B}}}{R_{\text{B}}}{d_{{\rm{2D}},m}}\left( {1 - v} \right)}}{\left( h - T_{\text{U}} \right)\pi }} \right)} 
	\end{array} \right)
	\end{aligned}
	\end{equation}

\begin{IEEEproof}
	See Appendix B.
\end{IEEEproof}

%


\textit{Remark 1}: In extremely high blockage density scenarios, since $E\left( {{\gamma _\varnothing}} \right)$ monotonically decreases with $\lambda_{\rm{B}}$ and $R_{\rm{B}}$, while monotonically increases with $v$, the expected received SNR at any UE location decreases with the density of blockages, the size of each blockage, and the loss caused by each blockage.

\textit{Remark 2}: In a given blockage scenario of an extremely high density and for a given number of IRSs, since $E\left( {{\gamma _\varnothing}} \right)$ monotonically increases with $h$, deploying the IRSs at a higher position results in a larger expected received SNR at any UE location. 

\textit{Remark 3}: In a given blockage scenario of an extremely high density and for a given number and deployment height of IRSs, the expected received SNR varies with the UE location.

\subsection{Expected FB Capacity} 
Based on (\ref{pro}) and the FB capacity under $\Xi$ in (21), the expected FB capacity for all possible $\Xi$ at a specific UE location is defined as
\begin{equation}\label{EFBC}
C^{\text{FB}}  = \sum\limits_\Xi ^{} {{\zeta _\Xi }E\left( {C_{ \Xi  }^{\text{FB}}} \right)}.
\end{equation}

Note that the expected FB capacity can also be used to evaluate the performance for eMBB services, that are usually evaluated using the Shannon capacity. This is because the Shannon capacity, ${\log _2}( {1 + \gamma _\Xi})$, is 
a special case of the FB capacity when the blocklength is infinite, and the difference between the former and the latter for a same channel is a constant determined by the decoding error probability and the blocklength, as shown in \cite[eq. (1)]{FBPoor}.

\textbf{Theorem 2}: For an extremely high blockage density (i.e., $\lambda_b$ is large but finite) and low transmit SNR, 
$C^{\text{FB}}$ approaches $ E\left( {C_\varnothing^{\text{FB}}} \right)$, and is upper bounded by
\begin{equation}
\begin{aligned}
\bar{C}&={\log_2}\left( 1 + E\left(\gamma _{ \varnothing}\right) \right) \\
&- \sqrt {\frac{1}{S} - \frac{1}{{S{{\left( 1 + E\left(\gamma _{ \varnothing } \right)\right)}^2}}}} \frac{{{Q^{ - 1}}( \varepsilon )}}{{\ln 2}},
\end{aligned}
\end{equation}
where $ {C_\varnothing^{\text{FB}}} =  {C_\Xi^{\text{FB}}}$ for $\Xi$ being a null set, and $E\left(\gamma _{ \varnothing } \right)$ is given in (\ref{EMh}).
\begin{IEEEproof}
Since ${C_{ \Xi }^{\text{FB}}} $ is a concave function of $\gamma _{ \Xi }$,  based on the Jensen's inequality and (21), for $\Xi=\varnothing$, we have
\begin{equation}\label{jensen}
\begin{aligned}
E\left( {C_{ \varnothing }^{\text{FB}}} \right) &\le{\log_2}\left( 1 + E\left(\gamma _{ \varnothing}\right) \right) \\ &- \sqrt {\frac{1}{S} - \frac{1}{{S{{\left( 1 + E\left(\gamma _{ \varnothing } \right)\right)}^2}}}} \frac{{{Q^{ - 1}}( \varepsilon )}}{{\ln 2}}.
\end{aligned}
\end{equation}
\end{IEEEproof}
The upper bound given on the right-hand side of (\ref{jensen}) becomes tighter for a lower transmit SNR $\rho$. 

\begin{table}[htbp]
		\centering
	\caption{IRS deployment schemes}
	\begin{tabular}{|c|c|c|c|c|c|}
		\hline 
		$M$ &  1& 4&8 &12&16\\  
		\hline
		${n_{{L_1}}}$ or ${n_{{L_2}}}$& 0 &0 & 1 & 2 &3\\
		\hline
		$n_W$ & 1 & 4 & 6 & 8& 10\\
		\hline
		$N_{\text{h}}$ & 32 & 16 & 12 & 10 &10\\
		\hline 
		$N_{\text{v}}$ & 30 & 15 & 10 & 8 &6\\
		\hline 
	\end{tabular}
	\label{table}
		\vspace{10pt}
	\centering
	\caption{Main simulation assumptions}
	\begin{tabular}{|c|c|}
		\hline 
		\textbf{Parameter name}  & \textbf{Parameter value}\\  
		\hline 
		Frequency (GHz) & 28 \\ 
		\hline
		Factory width $W$ (m)& 50\\
		\hline
		Factory length $L$ (m)& 40\\
		\hline
		Factory height $T_{\text{F}}$ (m) & 5 \\
		\hline
		Inter IRS element spacing $l=\mu/2$ (m)& 0.0054\\
		\hline
	    The $x$ coordinate of the fixed shelf $X_{\text{U}}$ (m) & 19.5\\
		\hline 
		Number of total IRS elements ${N}$ & 960 \\ 
		\hline 
		The power loss of the fixed shelf $\omega$ (dB)  & 20 \\ 
		\hline
		The power loss of a random blockage $\upsilon$ (dB) \cite{blockloss} & 20 \\
		\hline
		The width of a random blockage $R_{\text{B}}$ (m) \cite{3gpp} & 2.5\\
		\hline
		The maximum height of the blockages $T_{\text{B}}$ (m) \cite{3gpp} & 1.7 \\
		\hline
		UE height $T_{\text{U}}$ (m) & 0.5\\
		\hline
		Transmit power $P_{\text{T}}$ (dBm) & 22\\
		\hline
		Noise floor $\sigma$ (dB) & 9\\
		\hline
		Bandwidth $Z$ (MHz) & 400\\
		\hline
		Transmit antenna gain $G_{\text{T}}$ (dBi) & 24\\
		\hline
		Receive antenna gain $G_{\text{R}}$ (dBi) & 10\\
		\hline
		Capacity threshold $R$ (bit/s/Hz) & 0.1\\
		\hline
		Blocklength $S$ (nat) & 200\\
		\hline
		Decoding error rate $\epsilon$ & 1E-9\\
		\hline
	\end{tabular}
	\label{table}
\end{table}

\subsection{Expected Outage Probability}
Based on (\ref{pro}) and the outage probability under $\Xi$ in (22), the expected outage probability for all possible $\Xi$ at a specific UE location is defined as
\begin{equation}\label{EOP}
P = \sum\limits_\Xi ^{} {{\zeta _\Xi } P_{ \Xi  }  }.
\end{equation}

The closed-form expression of (\ref{EOP}) is hard to get because the probability density function (PDF) of the squared absolute value of channel ${{f}_\Xi}$ is analytically intractable due to ${f}_{\Xi}$ comprising of power loss caused by a random number of blockages multiplied by channel fading. 
Fortunately, it is easy to calculate \eqref{EOP} numerically, hence $P$ can be used to quickly evaluate the performance of various collocated and distributed IRS deployment.

\section{Numerical Results}
In this section, we consider the service area of $19.5 \times 50$ m$^2$ behind the shelf in a cubic factory of $40 \times 50 \times 5$ m$^3$, where we will investigate the collocated and distributed IRS deployment schemes by taking the number of IRSs being 1, 4, 8, 12, 16 as examples. The corresponding deployment configurations are shown in Table I.
The area of interest is sampled at 2 m resolution, resulting in 250 sampling UE locations over the service area.
For every UE location, the Monte Carlo simulation results of the three metrics, i.e., the expected received SNR, the expected FB capacity, and the expected outage probability, are calculated averaged over 1E7 realisations, comprised of 2500 random blockage drops times 4000 channel fading coefficients. 
Other parameters used in the simulations are given in Table \ref{table}.
Note that for URLLC services, low expected outage probability and high expected FB capacity are desired, while for eMBB services, high expected received SNR and high expected FB capacity are desired.

\begin{figure} [t]
	\centering
	\includegraphics[width=2.9in]{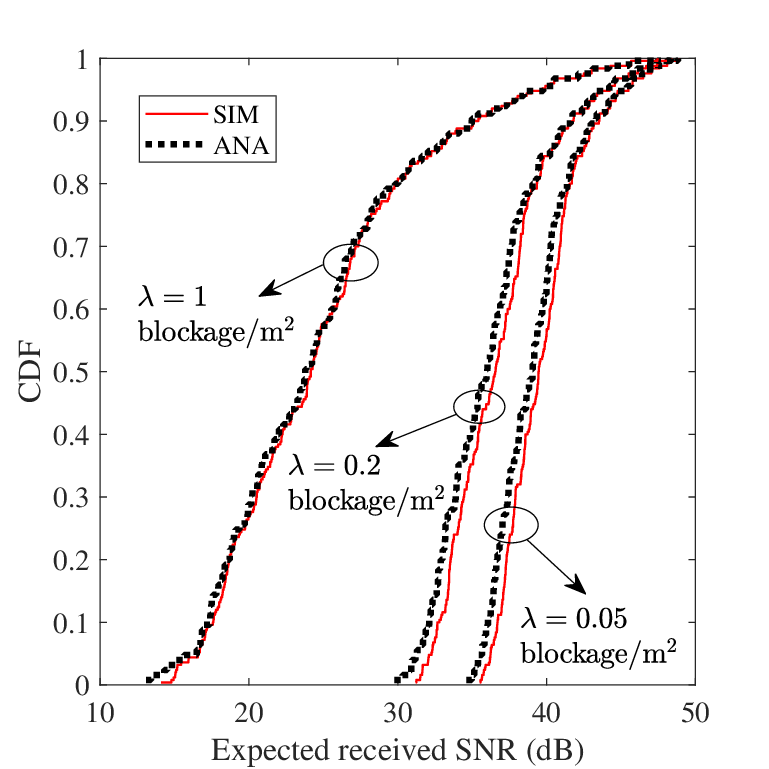}
	\caption{The CDF of the expected received SNRs at 250 sampling UE locations over the service area of $19.5 \times 50$ m$^2$ for $\lambda_{\text{B}}$ being 0.05, 0.2, 1 blockage/m$^2$, where $M=8$, $P_{\rm{T}}=30$ dBm, and $h=4$ m. Lines represent the simulation values while markers represent the analytical values calculated using (24).}
	\centering
	\includegraphics[width=2.9in]{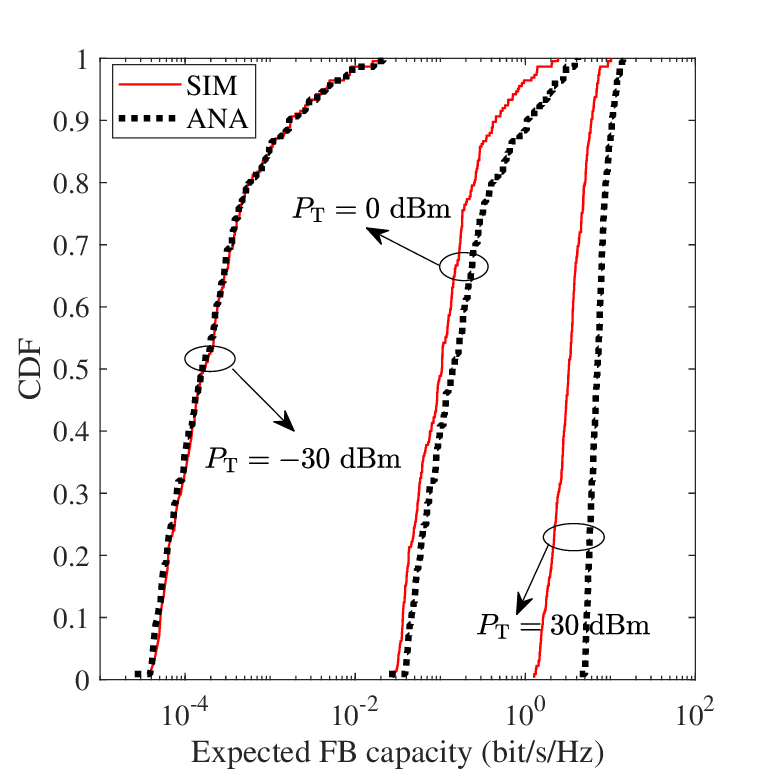}
	\caption{The CDF of the expected FB capacities at 250 sampling UE locations over the service area of $19.5 \times 50$ m$^2$ for $P_{\rm{T}}$ being -30, 0 30 dBm, where $M=12$, $\lambda_{\text{B}}=1$ blockage/m$^2$, and $h=4$ m. Lines represent the simulation values while markers represent the analytical values calculated using (26).}
\end{figure}

Fig. 2 shows the CDF of the expected received SNR for $\lambda_{\text{B}}$ being 0.05, 0.2, 1 blockage/m$^2$, where $M=8$, $P_{\rm{T}}=30$ dBm, and $h=4$ m. As the blockage density increases, the CDF lines of the expected received SNRs move left, leading to lower received SNRs at a relatively low percentile, which confirms \textit{Remark 1}. This is intuitive because the UE located further away from the BS and/or IRSs would be affected more significantly by the blockage density. Besides, the analytical results in (24) become tighter to the simulation results when the blockage density gets higher, which validates \textit{Theorem 1}. 

Fig. 3 depicts the CDF of the expected FB capacity for $P_{\rm{T}}$ being 30 dBm, 0 dBm, and -30 dBm, where $M=12$, $\lambda_{\text{B}}=1$ blockage/m$^2$ and $h=4$ m. We can see that, the expected FB capacity decreases when the transmit power is reduced. Meanwhile, by reducing the transmit power, the gap between the simulation and analytical results calculated using (26) gets smaller, which verifies \textit{Theorem 2}.

\begin{figure} [t]
	\begin{minipage}[t]{0.32\textwidth}
	\centering
	\includegraphics[width=2.3in]{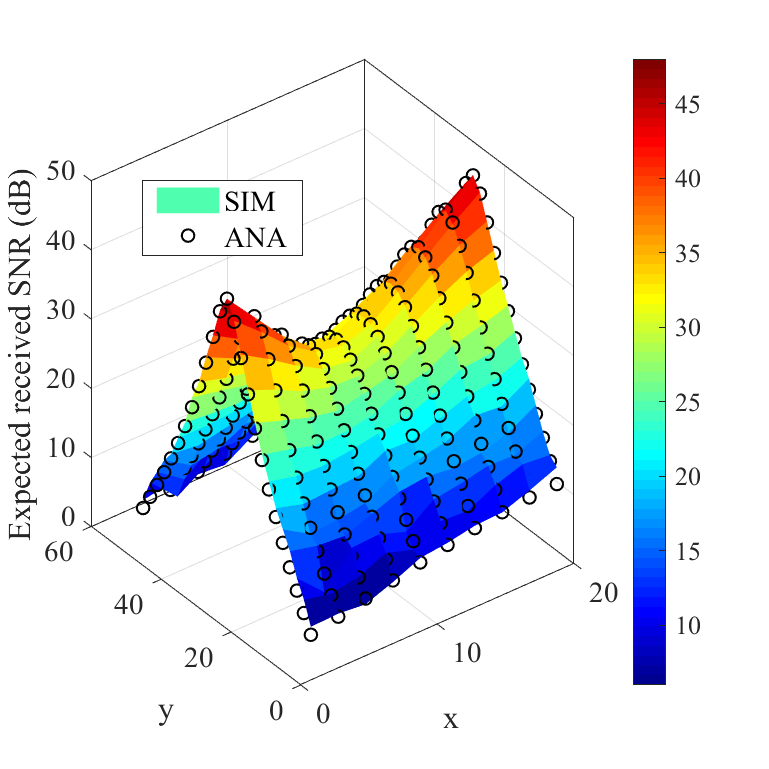}\\
	\centering
	\footnotesize(a) $M=1$\\
\end{minipage} 
	\begin{minipage}[t]{0.32\textwidth}
	\includegraphics[width=2.3in]{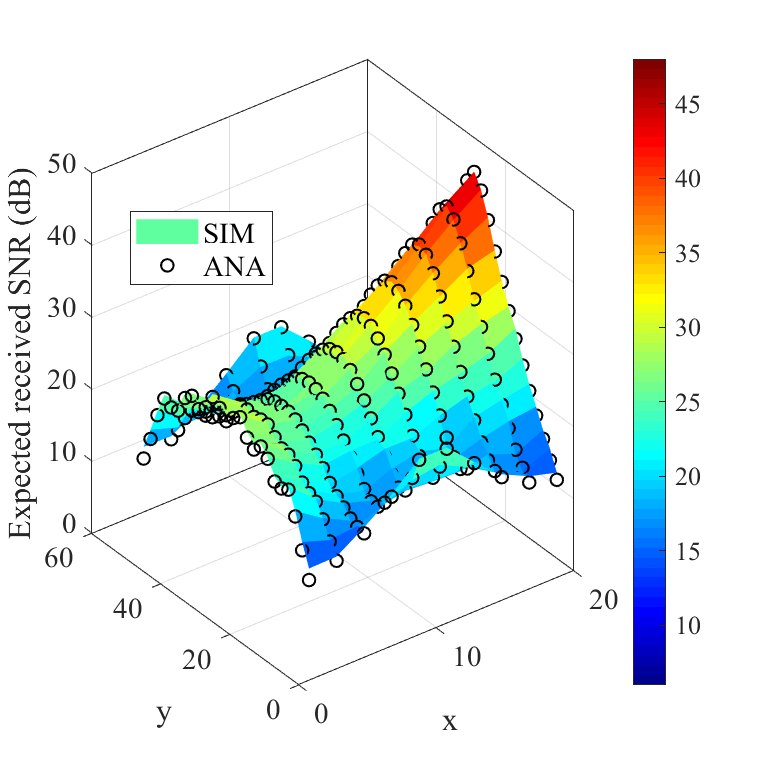}\\
	\centering
	\footnotesize(b) $M=8$\\
\end{minipage}
	\begin{minipage}[t]{0.32\textwidth}
	\includegraphics[width=2.3in]{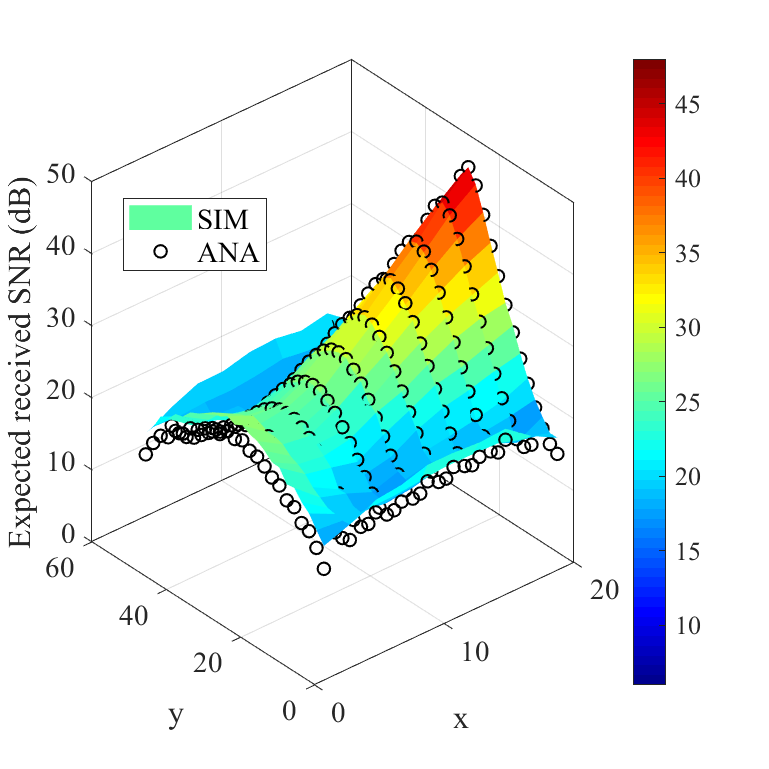}\\
	\centering
	\footnotesize(c) $M=16$\\ 
\end{minipage}
	\caption{The expected received SNRs at 250 sampling UE locations over the service area of $19.5 \times 50$ m$^2$ for (a) $M=1$, (b) $M=8$, (c) $M=16$, where $\lambda_{\text{B}}=1$ blockage/m$^2$, $P_{\text{T}}=30$ dBm, and $h=4$ m. Surfaces represent the simulation values while markers represent the analytical values obtained using (24).}
\end{figure}

\begin{figure} [htbp]
	\centering
	\includegraphics[width=3in]{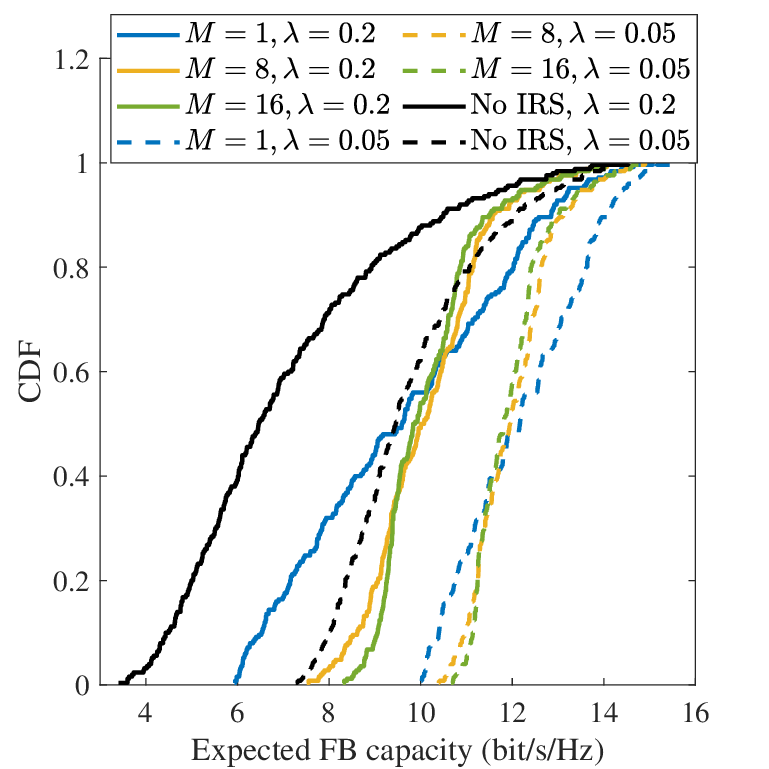}
	\caption{The CDF of the expected FB capacities at 250 sampling UE locations over the service area of $19.5 \times 50$ m$^2$ for $M$ being 1, 8, 16, where $\lambda_{\text{B}}$ is either 0.05 or 0.2 blockage/m$^2$, and $h=4$ m.}
\end{figure}

Figs. 4(a)-(c) present the expected received SNR for $M$ being 1, 8 and 16, respectively, where $\lambda_{\text{B}}=1$ blockage/m$^2$, $P_{\text{T}}=30$ dBm, and $h=4$ m. 
We observe that the fairness of the expected received SNRs over the service area is substantially enhanced.
The expected received SNRs of the UE located close to the corners or some walls of the room can be improved as the number of IRSs increases. When the number of IRSs increases from 1 to 8 and 16, the UE at the worst location that receives minimum expected received SNRs can receive 7.5 and 10.7 more dB, respectively. However, comparing the 1-IRS deployment and 16-IRS deployment, the UEs close to the BS or the IRS(s) have to make some compromise, as evidenced by the expected received SNR of the UE located at (1, 25, 0.5) and (20, 25, 0.5) decreased by 18.4 dB and 0.05 dB, respectively. 
Note that most of the analytical results calculated by (24) match well with the simulation results, while the gap between analytical and simulation results gets larger as the number of IRSs increases, especially for the UE locations near the 3 surrounding walls. That is because the number of blockages intersecting each IRS-UE link is not strictly independent to each other as the UE gets closer to the IRSs and the space between adjacent IRSs becomes smaller. Nonetheless, the analytical results calculated by (24) show the same trend as the simulated results of the expected received SNRs at all UE locations, and hence can be used to quickly evaluate and analyse the IRS deployment in smart factories \cite{blockage}.


%
%
%

In Fig. 5, the CDF of the expected FB capacity for $M$ being 1, 8, 16 under  $\lambda_{\text{B}}=0.2$ or 0.05 blockage/m$^2$ scenarios are presented, where $P_{\rm{T}}=30$ dBm, and $h=4$ m. The no-IRS deployment schemes are also plotted as benchmarks. For a same blockage density, at relatively low percentiles, the expected FB capacity increases with the number of IRSs. The CDFs of the expected FB capacity converge for large numbers of IRSs, e.g., $M = 8$ and $M = 16$. This indicates that the expected FB capacity stops further increasing when the number of IRSs is sufficiently large, especially at relatively high percentiles. For instance, comparing the 16-IRS scheme and the 1-IRS scheme with the no-IRS scheme, respectively, the minimum capacities are raised by 75\% and 147\%, when $\lambda_{\text{B}}=0.2$ blockage/m$^2$. Under the same comparison, for $\lambda_{\text{B}}=0.05$ blockage/m$^2$ scenarios, the minimum capacities are upgraded by 38\% and 46\%, respectively.
With the increase of blockage density, the expected FB capacity decreases.
\begin{figure} [t]
	\begin{minipage}[t]{0.5\textwidth}
		\centering
		\includegraphics[width=3in]{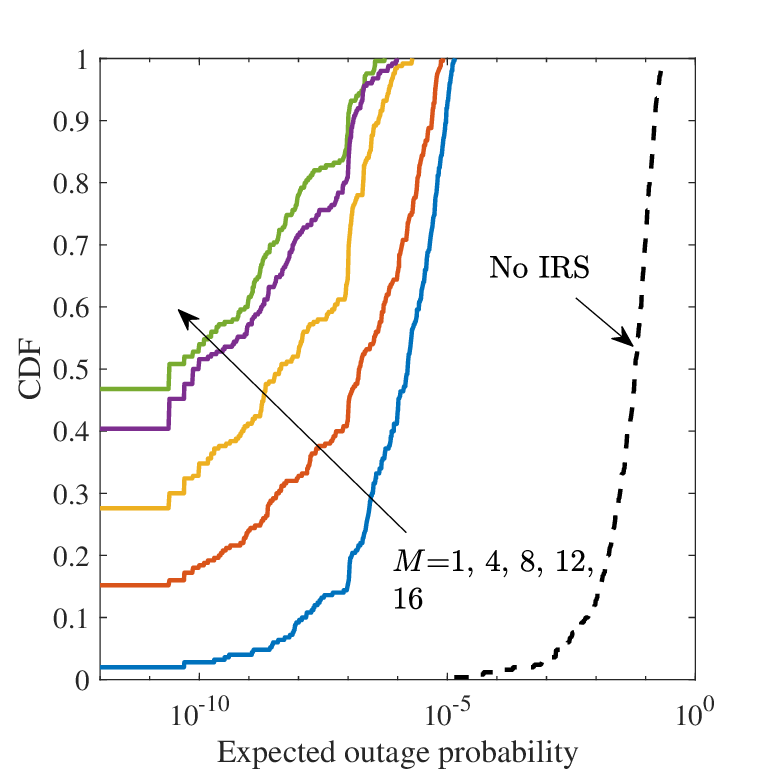}\\
		\centering
		\footnotesize (a) $\lambda_{\text{B}}=0.2$ blockage/m$^2$\\
	\end{minipage}
	\vspace{0mm}
	\begin{minipage}[t]{0.5\textwidth}
		\includegraphics[width=3in]{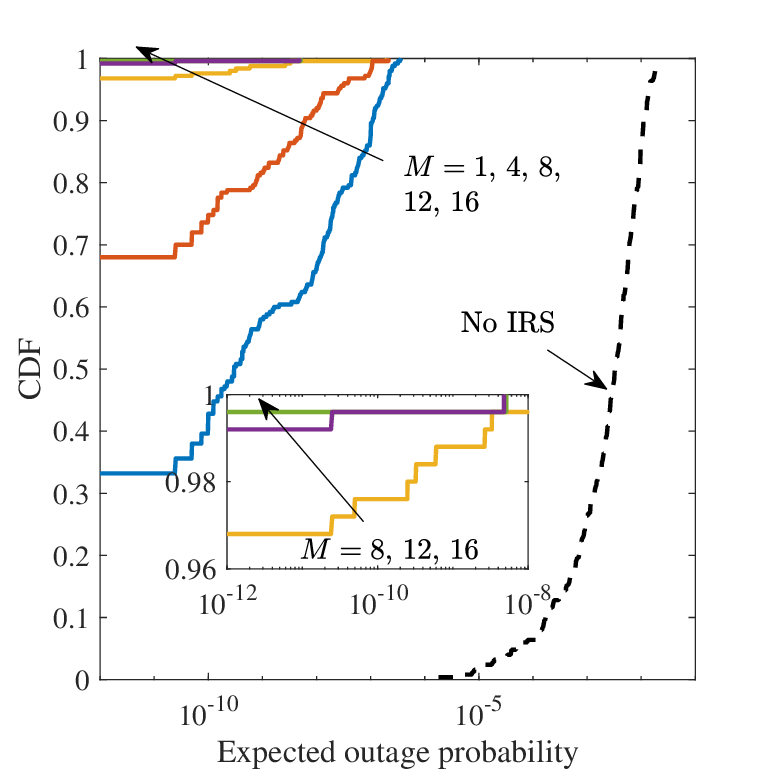}\\
		\centering
		\footnotesize (b) $\lambda_{\text{B}}=0.05$ blockage/m$^2$\\
	\end{minipage}
	\caption{The CDF of the expected outage probability at 250 sampling UE locations over the service area of $19.5 \times 50$ m$^2$ for $M$ being 1, 4, 8, 12, 16 under (a) $\lambda_{\text{B}}=0.2$ blockage/m$^2$, (b) $\lambda_{\text{B}}=0.05$ blockage/m$^2$ scenarios, where $P_{\text{T}}=30$ dBm, and $h=4$ m.}
\end{figure}

Figs. 6(a)-(b) illustrate the CDF of the expected outage probability for $M$ being 1, 4, 8, 12, 16, in $\lambda_{\text{B}}=0.2$ blockage/m$^2$ and $\lambda_{\text{B}}=0.05$ blockage/m$^2$ scenarios, respectively, with the no-IRS schemes plotted as benchmarks, where $P_{\rm{T}}=30$ dBm, and $h=4$ m.
For both considered blockage densities, compared with the no-IRS deployment scheme, the expected outage probability can be depressed by at least four orders of magnitude with the aid of IRS deployment.
When deploying distributed IRSs, the expected outage probability can be further reduced by one or two orders of magnitude, but the reduction slows down when the number of IRSs exceeds 12.
Comparing Fig. 6(a) with Fig. 6(b), we can see that the minimum required number of IRSs for achieving the same expected outage probability increases with the blockage density.

To summarise, in terms of the expected received SNR and expected FB capacity, leveraging more distributed IRSs to assist wireless communications will improve the fairness among all UE locations by sacrificing the superior performance of a small group of UEs enabled by collocated IRSs. In addition, the expected outage probability can also be suppressed by using distributed IRSs.


\subsection{The impact of $M$ and $h$ for practical high blockage density scenario (i.e., $\lambda_{\text{B}}=0.2$ blockage/m$^2$)}
In Fig. 7, the mean/minimum/maximum values of the expected received SNR, the expected FB capacity, and the expected outage probability among 250 sampling UE locations versus the number of IRSs for different IRS height of 2, 3, 4 m are shown, where $\lambda_{\text{B}}=0.2$ blockage/m$^2$, $P_{\text{T}}=30$ dBm. 

Generally, under any number of IRSs deployed, raising the height of IRSs will always enhance the average/minimum expected received SNR and average/minimum expected FB capacity, and reduce the average/maximum expected outage probability, which demonstrates that higher deployment of IRS ensures greater availability of the IRS-UE links, as shown in \textit{Remark 2}.

\begin{figure*}[t] 
	\begin{minipage}[t]{0.31\textwidth}
		\includegraphics[width=2.3in]{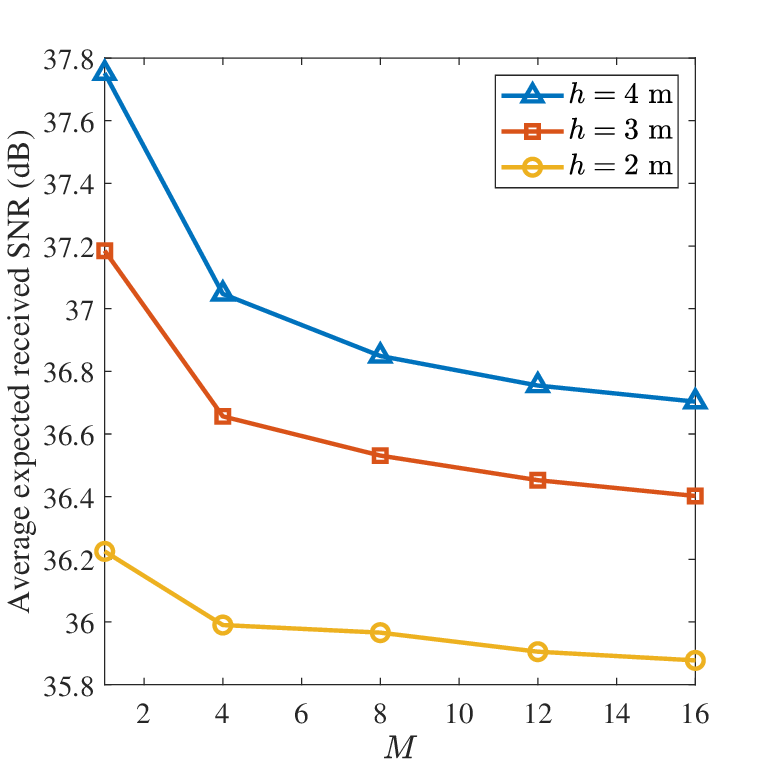}
		\centering
		\footnotesize(a) 
		\label{SNRcapacity}
		\vspace{0pt}
	\end{minipage}
	\hspace{3mm}
	\begin{minipage}[t]{0.31\textwidth}
		\includegraphics[width=2.3in]{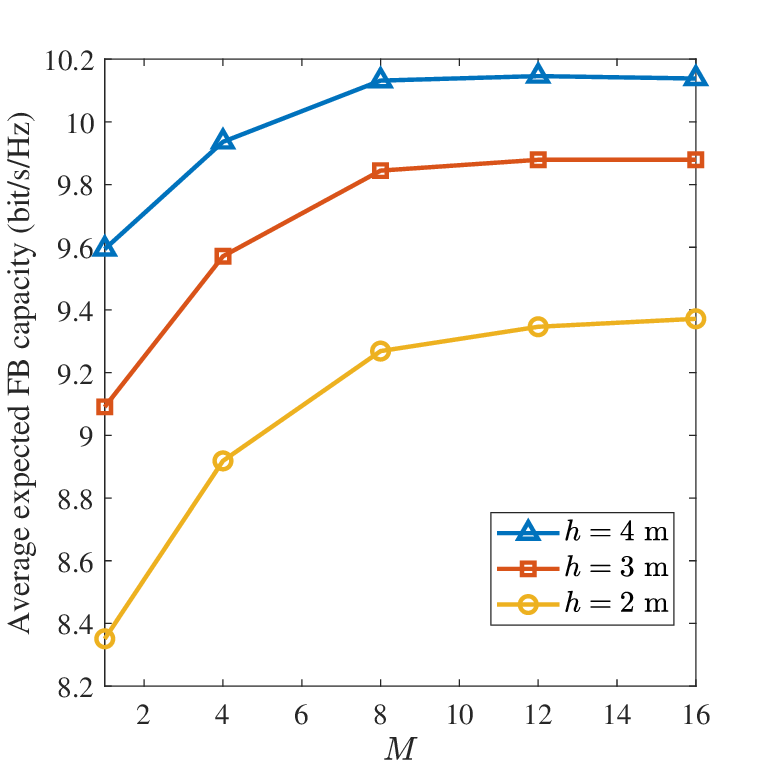}
		\centering
		\footnotesize(b) 
		\label{eigendis}
		\vspace{0pt}
	\end{minipage}
	\hspace{3mm}
	\begin{minipage}[t]{0.31\textwidth}
		\includegraphics[width=2.3in]{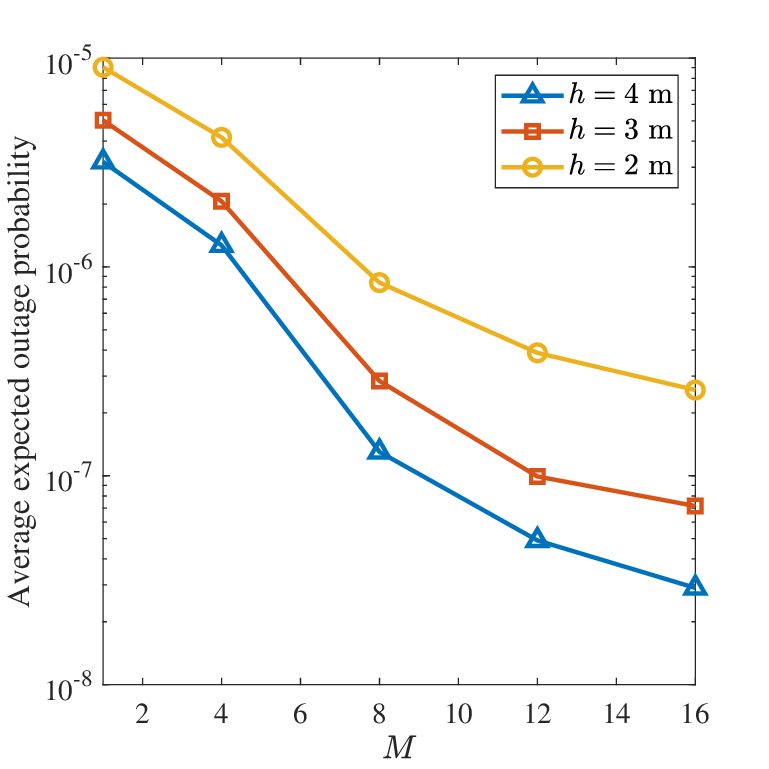}
		\centering
		\footnotesize(c) 
		\label{eigendis}
		\vspace{0pt}
	\end{minipage}
	\hspace{3mm}
	\label{TR}
	\vspace{0mm}
	\vspace{-0pt}
	\begin{minipage}[t]{0.31\textwidth}
		\includegraphics[width=2.3in]{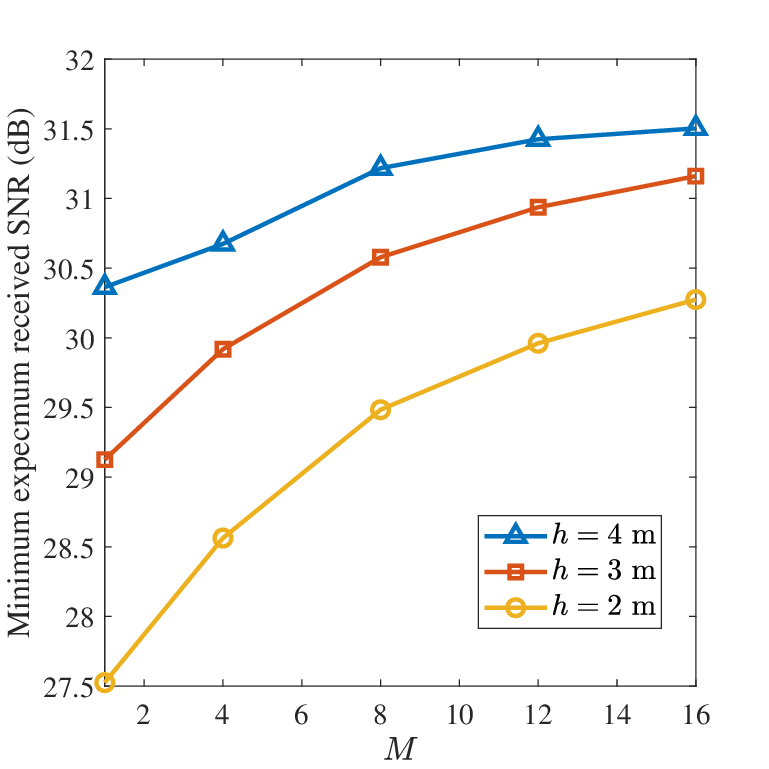}
		\centering
		\footnotesize(d) 
		\label{SNRcapacity}
		\vspace{0pt}
	\end{minipage}
	\hspace{3mm}
	\begin{minipage}[t]{0.31\textwidth}
		\includegraphics[width=2.3in]{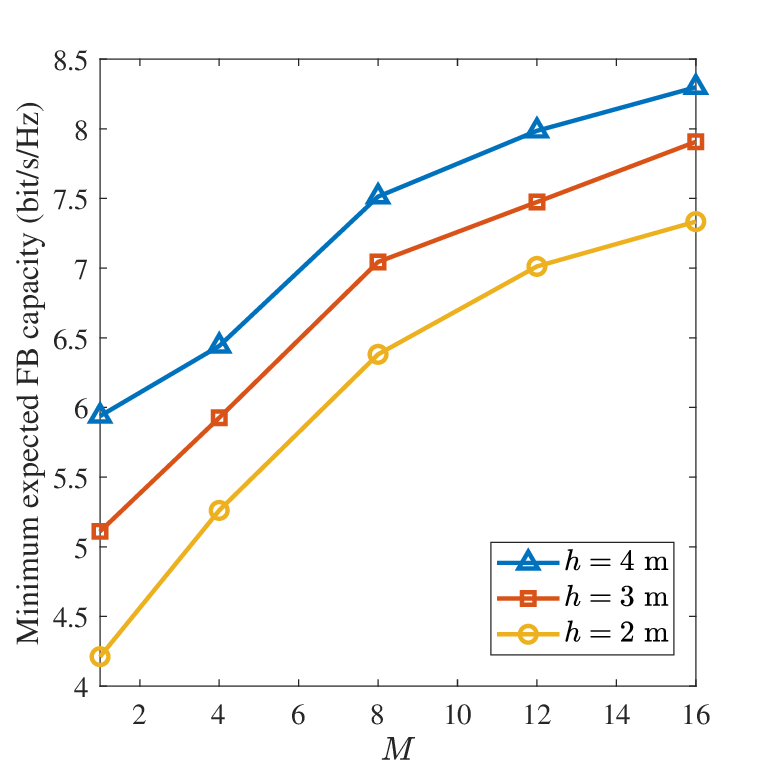}
		\centering
		\footnotesize(e) 
		\label{eigendis}
		\vspace{0pt}
	\end{minipage}
	\hspace{3mm}
	\begin{minipage}[t]{0.31\textwidth}
		\includegraphics[width=2.3in]{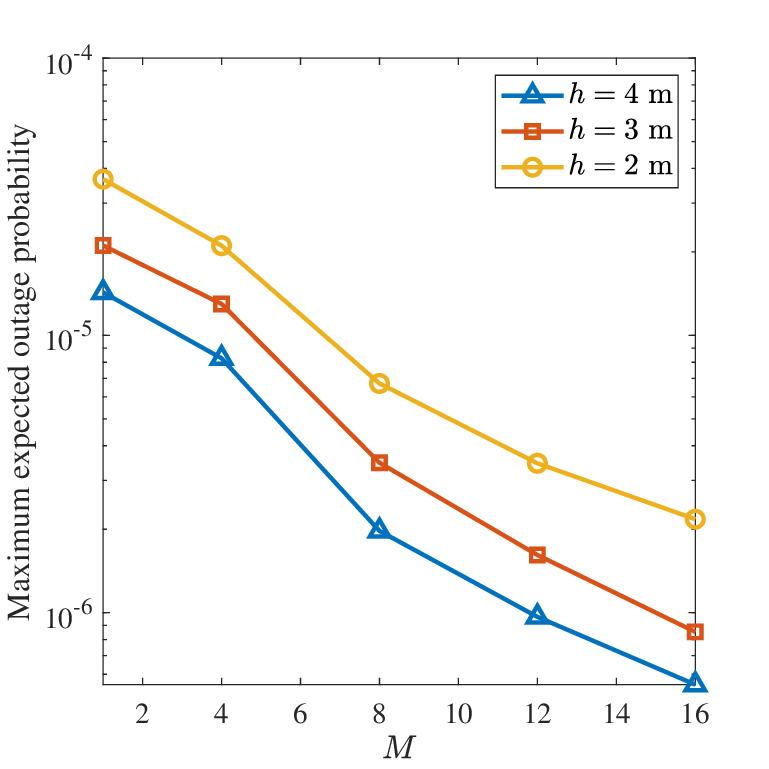}
		\centering
		\footnotesize(f) 
		\label{eigendis}
		\vspace{0pt}
	\end{minipage}
	\hspace{3mm}
	\label{TR}
	\vspace{-0mm}
	\caption{The mean/minimum/maximum values of the expected received SNR, expected FB capacity, and expected outage probability among 250 sampling UE locations versus the number of IRSs for different IRS height of 2, 3, 4 m, where $\lambda_{\text{B}}=0.2$ blockage/m$^2$, $P_{\text{T}}=30$ dBm.}
\end{figure*}
With regards to the expected received SNR, we can tell from in Fig. 7(a) and Fig. 7(d) that its average decreases with the number of IRSs, however, its minimum increases with the number of IRSs. Compared with 16-IRS deployment, the 1-IRS deployment trades a 1.05 dB decrease in average performance for a 1.14 dB improvement in minimum performance when the IRS is deployed at 4 m high, and trades a 0.35 dB decline in average performance for a 3.98 dB enhancement in minimum performance when the IRS is deployed at 2 m high. That is to say, even if there is limit in the height of the IRS deployment, decentralising the IRS elements into more units benefits the fairness of received SNRs among different UE locations.

Fig. 7(b) and Fig. 7(e) show the average and minimum expected FB capacity with respect to the number of IRSs, respectively. We can see that both the mean and minimum expected FB capacity increase with a declining growth speed when more distributed IRSs are deployed. 
When enlarging the IRS units from 1 to 16, the minimum expected FB capacity can be enhanced nearly 71\%, 52\% and 38\%, respectively, for the IRS height of 2, 3, 4 m.
Although the average performance converges to a constant when the number of IRSs exceeds 8, the minimum performance at the worst UE location can still achieve nearly 1 more bit/s/Hz when continue increasing the number of IRSs to 16.

Fig. 7(c) and Fig. 7(f) illustrate the average and maximum expected outage probability versus the number of IRSs, respectively, where all the curves show a downward trend against the number of IRSs. Comparing the 16-IRS scheme with the 1-IRS scheme, the average expected outage probability can be reduced by a factor of 110, 70, and 35, for the IRS deployment height of 4 m, 3 m, and 2 m, respectively, while the maximum expected outage probability can be reduced by a factor of 26, 25, 17, for the IRS deployment height of 4 m, 3 m, and 2 m, respectively. Obviously, using distributed IRS deployment schemes will pull up the probability of meeting the capacity requirement, allowing the wireless system to support more demanding use cases. It is noteworthy that deploying IRSs higher on the chosen walls make the reduction in expected outage probability more pronounced.

Overall, in order to combat the heavy blockage effects in high blockage density scenarios, distributing the fixed number of IRS elements into a larger amount of IRSs and deploying them higher on the chosen walls is lucrative for URLLC services.
On the contrary, for eMBB services, it is not advisable to deploy a large number of distributed IRSs, e.g., increasing the number of IRSs beyond 8 offers trivial gains.

\begin{figure*}[t] 
	\begin{minipage}[t]{0.31\textwidth}
		\includegraphics[width=2.3in]{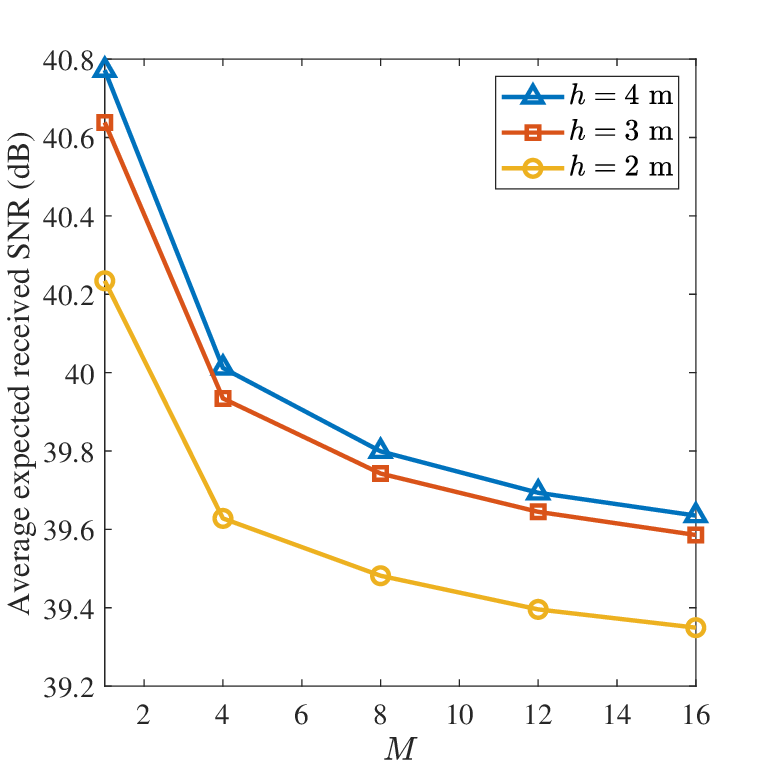}
		\centering
		\footnotesize(a) 
		\label{SNRcapacity}
		\vspace{0pt}
	\end{minipage}
	\hspace{3mm}
	\begin{minipage}[t]{0.31\textwidth}
		\includegraphics[width=2.3in]{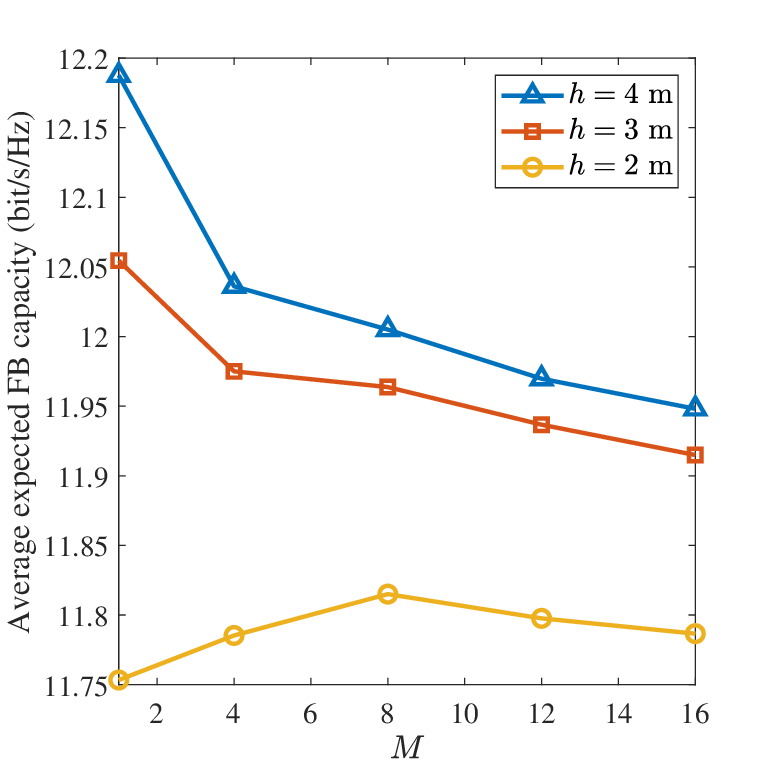}
		\centering
		\footnotesize(b) 
		\label{eigendis}
		\vspace{0pt}
	\end{minipage}
	\hspace{3mm}
	\begin{minipage}[t]{0.31\textwidth}
		\includegraphics[width=2.3in]{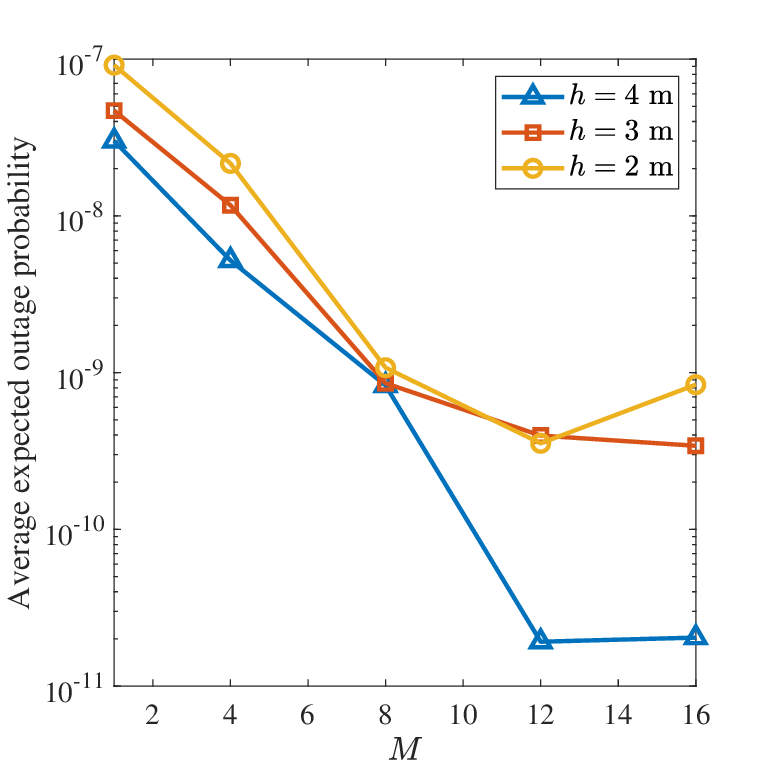}
		\centering
		\footnotesize(c) 
		\label{eigendis}
		\vspace{0pt}
	\end{minipage}
	\hspace{3mm}
	\label{TR}
	\vspace{0mm}
	\begin{minipage}[t]{0.31\textwidth}
		\includegraphics[width=2.3in]{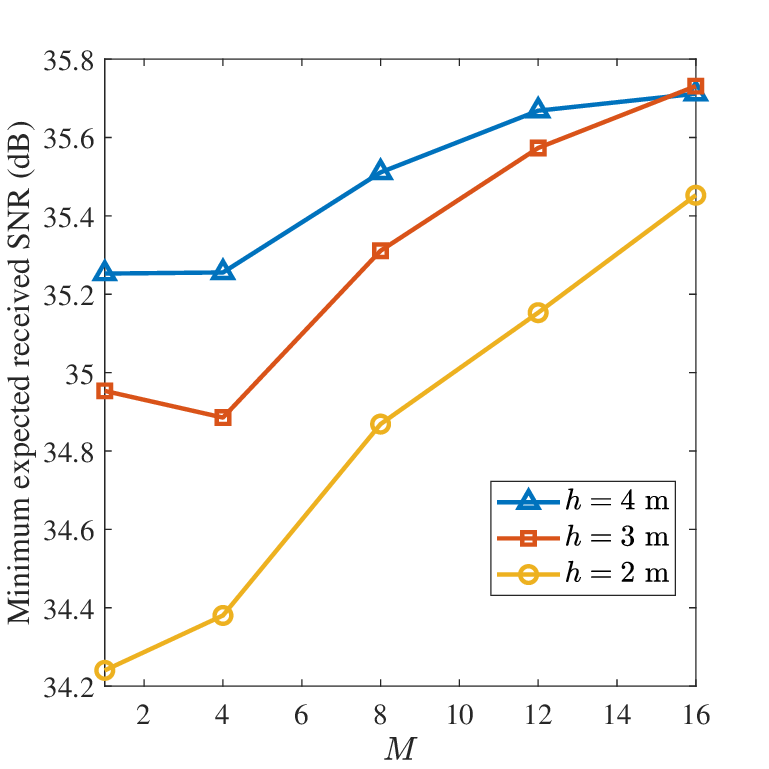}
		\centering
		\footnotesize(d) 
		\label{SNRcapacity}
		\vspace{0pt}
	\end{minipage}
	\hspace{3mm}
	\begin{minipage}[t]{0.31\textwidth}
		\includegraphics[width=2.3in]{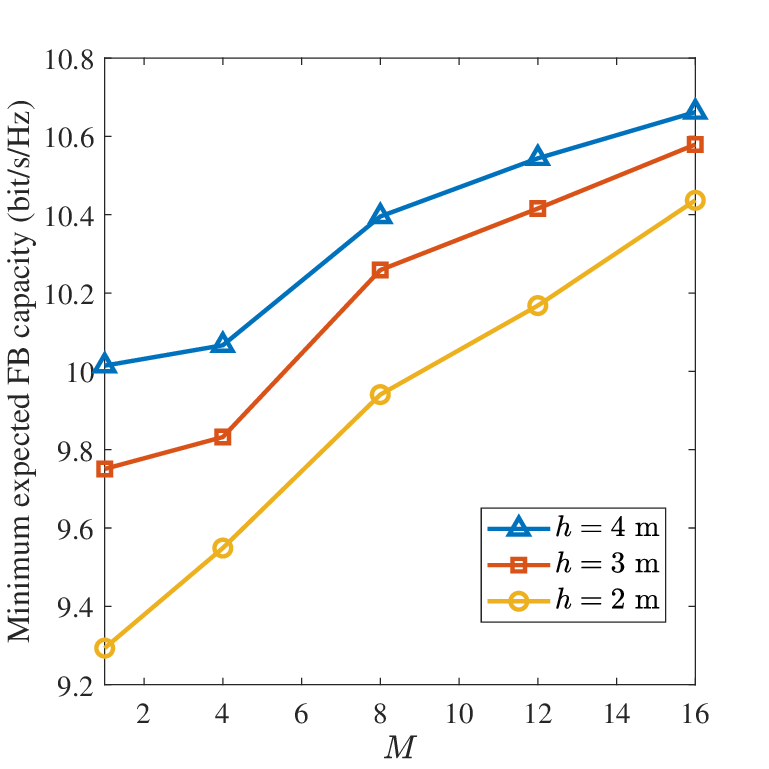}
		\centering
		\footnotesize(e) 
		\label{eigendis}
		\vspace{0pt}
	\end{minipage}
	\hspace{3mm}
	\begin{minipage}[t]{0.31\textwidth}
		\includegraphics[width=2.3in]{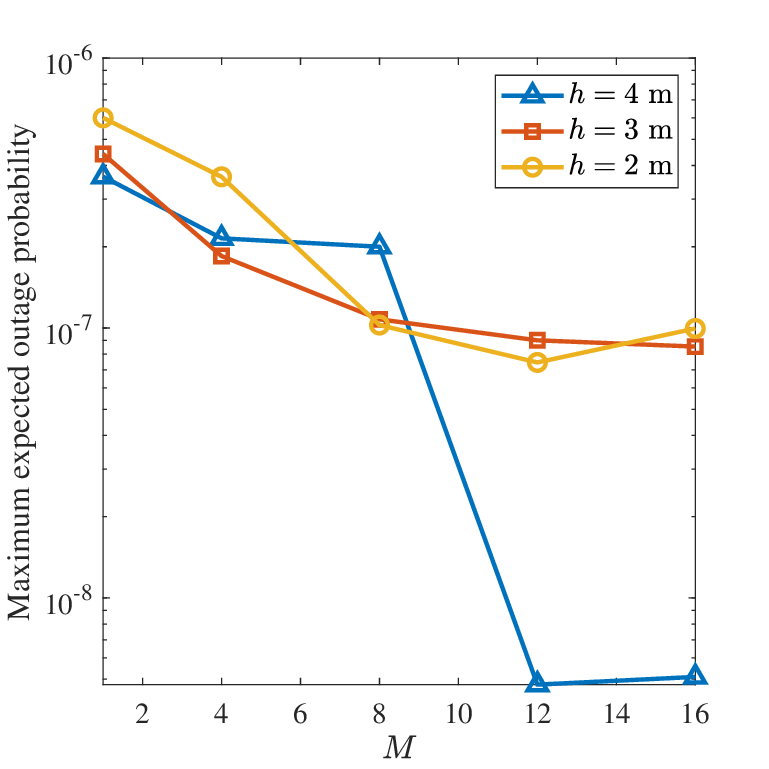}
		\centering
		\footnotesize(f) 
		\label{eigendis}
		\vspace{0pt}
	\end{minipage}
	\hspace{3mm}
	\label{TR}
	\vspace{-0mm}
	\caption{The mean/minimum/maximum values of the expected received SNR, expected FB capacity, and expected outage probability among 250 sampling UE locations versus the number of IRSs for different IRS height of 2, 3, 4 m, where $\lambda_{\text{B}}=0.05$ blockage/m$^2$, $P_{\text{T}}=30$ dBm.}
\end{figure*}

\subsection{The impact of $M$ and $h$ for practical low blockage density scenario (i.e., $\lambda_{\text{B}}=0.05$ blockage/m$^2$)}

In Fig. 8, the mean/minimum/maximum values of the expected received SNR, the expected FB capacity, and the expected outage probability among 250 sampling UE locations versus the number of IRSs for different IRS height of 2, 3, 4 m are depicted, where $\lambda_{\text{B}}=0.05$ blockage/m$^2$, $P_{\text{T}}=30$ dBm.

As shown in Fig. 8(a) and Fig. 8(b), the average expected received SNR and the averaged expected FB capacity basically decrease with the number of IRSs, except the values for the average expected FB capacity for the IRS height of 2 m remain almost unchanged.
As shown in Fig. 8(d) and Fig. 8(e), the minimum expected received SNR and the minimum expected FB capacity show slight upward trends when the number of IRSs is enlarged.

As mentioned before, the minimum performance can be improved by sacrificing a little average performance. To illustrate, adding up the number of IRSs at 2 m high from 1 to 16 causes 0.89 dB decrease in averaged expected received SNR and 0.03 bit/s/Hz increase in averaged expected FB capacity, while leading to a 1.13 dB increase in minimum expected received SNR and a 1.14 bit/s/Hz increase in minimum expected FB capacity. For the IRS height being 4 m, deploying 16 IRSs instead of 1 IRS compromises the average expected received SNR of 1.14 dB and the average expected FB capacity of 0.24 bit/s/Hz, but at the same time brings about a 0.46 dB increase in minimum expected received SNR and a 0.65 bit/s/Hz increase in minimum expected FB capacity. Moreover, we observe that the improvement in the minimum expected SNR and minimum expected capacity becomes more limited when the IRSs are lifted from 3 m to 4 m high than from 2 m to 3 m, especially for the number of IRSs being 12 and 16.

In Fig. 8(c) and Fig. 8(f), the average and maximum expected outage probability are generally getting lower when the amount of IRS grows from 1 to 16. The IRS deployment at 2 m and 3 m height show similar performance in the average and maximum expected outage probability for the amount of IRS being 1, 4, 8, 12, 16. In addition, for the number of IRSs being 1, 4, and 8, raising the IRS(s) to 4 m height does not change much the average/maximum expected outage probability as compared to the IRS deployment at 2 m or 3 m height, nonetheless, when the number of IRSs goes up from 8 to 12 or 16, the averaged/maximum expected outage probability will be significantly declined by more than 40 times.

Therefore, for practical low blockage densities, there is not much difference between the collocated and distributed IRS deployment schemes for eMBB services.
Nevertheless, for URLLC services, it is preferable to use more than 8 distributed IRSs deployed at 4 m high on the chosen walls to suppress the maximum expected outage probability. Moreover, it may not be very cost-effective to adopt more than 12 distributed IRSs for user fairness, due to the marginal gains in minimum expected FB capacity and maximum expected outage probability.


\section{Conclusions}
In this article, we have constructed a new channel model for analysing collocated or distributed IRS deployment in a cubic factory, taking the effects of interior blockages into account. Targeting eMBB and URLLC services in smart factories, we have proposed three performance metrics, i.e., the expected received SNR, the expected FB capacity, and the expected outage probability, where the expectation is taken over all possible blockage cases and channel fading realisations. 
For extremely high blockage densities, we have derived the closed-form expressions of the expected received SNR and expected FB capacity. The analytical expressions are validated by Monte Carlo simulations.
	
Based on our analytical and simulation results, we obtain the following insights into the deployment of multiple IRSs for supporting eMBB and URLLC services in smart factories.
\begin{itemize}
	\item For high blockage densities, distributing a fixed total number of IRS elements into more IRSs and deploying them higher on the chosen walls will lead to a higher expected received SNR and a higher expected FB capacity, thus benefiting both eMBB and URLLC services.
	\item For low blockage densities, eMBB services can be well supported by either collocated or distributed IRSs; while for URLLC services, deploying distributed IRSs higher on the chosen walls can effectively suppress the expected outage probability.
	\item In terms of the expected received SNR and expected FB capacity, leveraging more distributed IRSs will improve the fairness among all UE locations.
	\item The expected outage probability can be suppressed by using distributed IRSs. The minimum required number of IRSs for achieving the same expected outage probability increases with the blockage density.
    \item There exists an optimal number of distributed IRSs, beyond which the marginal gains in the expected received SNR, expected FB capacity, and expected outage probability become negligible.
    \item The expected received SNR or expected FB capacity or expected outage probability for the UEs located further away from the BS and/or IRSs would be affected more significantly by the blockage density, compared with the UEs located near the BS and/or IRSs.	
\end{itemize}

%
\newpage 
\section*{Appendix A}
Considering the height of a random blockage, the number of blockages that intersects the BS-UE link and the $m$th IRS-UE link are Poisson random variables, whose expectations can be respectively derived by \cite[Theorem 3]{blockage},
\begin{equation}\label{EB0}
E(B_{0})=\eta_1 E(B_{\text{2D},0}),
\end{equation} 
\begin{equation}\label{EBm}
E(B_{m})=\eta_2 E(B_{\text{2D},m}),
\end{equation}
where $B_{\text{2D},0}$ and $B_{\text{2D},m}$ denote
the number of blockages that intersects the projection lines in the XOY plane of the BS-UE link and the $m$th IRS-UE link, respectively, $\eta_1$ denotes the conditional probability of blocking the BS-UE link given its projection line in the XOY plane is intersected by a blockage, and $\eta_2$ denotes the conditional probability of blocking the IRS-UE link given its projection line in the XOY plane is intersected by a blockage.

Particulary, $B_{\text{2D},0}$ and $B_{\text{2D},m}$ are also Poisson random variables, whose expectations are given by \cite[Theorem 1]{blockage}
\begin{equation}\label{EB02D}
E(B_{\text{2D},0})=\frac{2{\lambda _{\text{B}}}{R_{\text{B}}}{d_{\text{2D},0}}}{\pi },
\end{equation} 
\begin{equation}\label{EBm2D}
E(B_{\text{2D},m})=\frac{2{\lambda _{\text{B}}}{R_{\text{B}}}{d_{\text{2D},m}}}{\pi }.
\end{equation} 

Since the height of the blockage is not greater than ${T_{\rm{B}}}$, the BS-UE link will not be blocked when a blockage is horizontally separated from the UE beyond $S_1$ m in the XOY plane, where
\begin{equation}
S_1 = \frac{{{T_{\rm{B}}} - {T_{\rm{U}}}}}{{{T_{\rm{F}}} - {T_{\rm{U}}}}}{d_{{\rm{2D}},0}}.
\end{equation}
Then
\begin{equation}{T_{s_1}} = \frac{{\left( {{T_{\rm{F}}} - {T_{\rm{U}}}} \right)s_1}}{{{d_{{\rm{2D}},0}}}} + {T_{\rm{U}}}
\end{equation}
represents the minimum height of a blockage intersecting the BS-UE link when the blockage is horizontally separated from the UE by $s_1$ m in the XOY plane, where $0 \le s_1 \le S_1$.
Hence, we obtain
\begin{equation}\label{eta1}
\eta_1= \frac{1}{{{d_{{\rm{2D}},0}}}}\int\limits_0^{S_1} {\left( {1 - \int\limits_{{T_{\rm{U}}}}^{{T_{s_1}}} {\frac{1}{{{T_{\rm{B}}} - {T_{\rm{U}}}}}{\rm{d}}t} } \right)} {\rm{d}}s_1
=\frac{{{T_{\rm{B}}} - {T_{\rm{U}}}}}{{2\left( {{T_{\rm{F}}} - {T_{\rm{U}}}} \right)}}.
\end{equation} 

Similarly, the $m$th IRS-UE link will not be blocked when a blockage is horizontally separated from the UE beyond $S_2$ m in the XOY plane, where 
\begin{equation}
S_2 = \frac{{{T_{\rm{B}}} - {T_{\rm{U}}}}}{h - {T_{\rm{U}}}}{d_{{\rm{2D}},m}}. 
\end{equation}
Then
\begin{equation}
{T_{s_2}} = \frac{{\left( {h - {T_{\rm{U}}}} \right)s_2}}{{{d_{{\rm{2D}},m}}}} + {T_{\rm{U}}}
\end{equation}
represents the minimum height of a blockage intersecting the IRS-UE link when the blockage is horizontally separated from the UE by $s_2$ m in the XOY plane, where $0 \le s_2 \le S_2$. Thus, we have
\begin{equation}\label{eta2}
\eta_2=\frac{1}{{{d_{{\rm{2D}},m}}}}\int\limits_0^{S_2} {\left( {1 - \int\limits_{{T_{\rm{U}}}}^{{T_{s_2}}} {\frac{1}{{{T_{\rm{B}}} - {T_{\rm{U}}}}}{\rm{d}}t} } \right)} {\rm{d}}s_2=\frac{{{T_{\rm{B}}} - {T_{\rm{U}}}}}{{2\left( {h - {T_{\rm{U}}}} \right)}}.
\end{equation}

By substituting (\ref{EB02D}) and (\ref{eta1}) into (\ref{EB0}) as well as substituting (\ref{EBm2D}) and (\ref{eta2}) into (\ref{EBm}), (\ref{blocker0}) and (\ref{blockerm}) are obtained.

Following \cite[Corollary 1]{blockage}, the LOS probability of the $m$th IRS-UE link is given in (\ref{losm}).

\section*{Appendix B}
The expected value of $\gamma_\Xi$ can be derived following (\ref{Err}) on the next page. For a general case $\Xi$ where $K_{{\rm{ru}},m} \neq 0$, $E\left( {\left| {{f_{{\rm{ru}},m,n}}} \right|} \right)$ is a function of $d_{m}$.
In addition, $E\left[ {{\upsilon ^{{B_{m,\Xi}}}}} \right]$ or $E\left[ {{\sqrt{\upsilon ^{{B_{m,\Xi}}}}}} \right]$ is also a function of $d_{m}$, as shown in Lemma 1. Hence, if $K_{{\rm{ru}},m} \neq 0$, then ${{\upsilon ^{{B_{m,\Xi}}}}}$ and ${{f_{{\rm{ru}},m,n}}}$ are not independent to each other.

However, in extremely high blockage density scenarios, (i.e., $\lambda_b$ is large but finite), all $M$ IRS-UE links are likely to be blocked, making $\Xi=\varnothing$,
${\zeta _{\varnothing }} \to 1$, ${K_{{\rm{ru}},m}} \to 0$, 
and ${{\bf{f}}_{{\rm{ru}},m}}$ follows Raleigh fading channel, for $m=1, 2, ..., M$,  with
$E\left( {\left| {{f_{{\rm{ru}},m,n}}} \right|^2} \right) =1$
and
$E\left( {\left| {{f_{{\rm{ru}},m,n}}} \right|} \right)=\frac{{\sqrt \pi  }}{2}$,
which means $E\left( {\left| {{f_{{\rm{ru}},m,n}}} \right|} \right)$ is no longer a function of $d_m$, and is independent of $E\left[ {{\upsilon ^{{B_{m}}}}} \right]$ or $E\left[ {{\sqrt{\upsilon ^{{B_{m}}}}}} \right]$, resulting in the decoupling of $E\left({{\sqrt{\upsilon ^{{B_{m}}}}}} {\left| {{f_{{\rm{ru}},m,n}}} \right|} \right)$ and $E\left({{\upsilon ^{{B_{m}}}}} {\left| {{f_{{\rm{ru}},m,n}}} \right|} \right)$ \cite[Corollary 5.2]{blockage}, thus leading to $\gamma \to E\left( {{\gamma _{\varnothing }}} \right)$, the expression after $\overset{(a)}{=}$ in (\ref{Erzero}).

\begin{equation}\setcounter{equation}{39}\label{Err}
\begin{array}{c}
\begin{aligned}
E\left( {{\gamma _\Xi }} \right)  &= \rho E\left( {{{\left( {\sqrt {{\beta _0}\omega {\upsilon ^{{B_{0}}}}} \left| {{f_{\rm{bu}}}} \right| + \sum\limits_{m = 1}^M {\sqrt {{\beta _m}{\upsilon ^{{B_{m,\Xi}}}}} \sum\limits_{n = 1}^{N/M} {\left| {{f_{{\rm{ru}},m,n}}} \right|} } } \right)}^2}} \right)\\ 
& = \rho\left( \begin{array}{l}
{\beta _0}\omega E\left[ {{\upsilon ^{{B_{0}}}}}  {{{\left| {{f_{\rm{bu}}}} \right|}^2}} \right]
\!+\! 2\sqrt {{\beta _0}\omega } \sum\limits_{m = 1}^M {\sqrt {{\beta _m}} E\left[ {\sqrt {{\upsilon ^{\left( {{B_{0}} \!+\! {B_{m,\Xi}}} \right)}}} \left| {{f_{\rm{bu}}}} \right|\sum\limits_{n = 1}^{N/M} {\left| {{f_{{\rm{ru}},m,n}}} \right|} } \right]} \\
\!+\! \sum\limits_{m = 1}^M {\sum\limits_{p = 1,p \ne m}^M {\sqrt {{\beta _m}{\beta _p}} E\left[ {\sqrt {{\upsilon ^{\left( {{B_{p,\Xi}} \!+\! {B_{m,\Xi}}} \right)}}} \sum\limits_{{n_1} = 1}^{N/M} {\left| {{f_{{\rm{ru}},p,{n_1}}}} \right|} \sum\limits_{{n_2} = 1}^{N/M} {\left| {{f_{{\rm{ru}},m,{n_2}}}} \right|} } \right]} }\\ 
\!+\! \sum\limits_{m = 1}^M {{\beta _m}E\left[ {{\upsilon ^{{B_{m,\Xi}}}}}  {{{\left( {\sum\limits_{n = 1}^{N/M} {\left| {{f_{{\rm{ru}},m,n}}} \right|} } \right)}^2}}\right]} 
\end{array} \right)\\
\end{aligned}
\end{array}
\end{equation}
\begin{equation}\setcounter{equation}{40}\label{Erzero}
\begin{array}{c}
\begin{aligned}
E\left( {{\gamma _{\varnothing}}} \right)&\overset{(a)}{=} \rho \left( \begin{array}{l}
{\beta _0}\omega E\left[ {{\upsilon ^{{B_{m}}}}} \right]E\left[ {{{\left| {{f_{\rm{bu}}}} \right|}^2}} \right] \\
\!+\! 2\sqrt {{\beta _0}\omega } \sum\limits_{m = 1}^M {\sqrt {{\beta _m}} E\left[ {\sqrt {{\upsilon ^{\left( {{B_0} + {B_m}} \right)}}} } \right]E\left[ {\sum\limits_{n = 1}^{N/M} {\left| {{f_{\rm{bu}}}} \right|\left| {{f_{{\rm{ru}},m,n}}} \right|} } \right]} \\
+ \sum\limits_{m = 1}^M {\sum\limits_{p = 1,p \ne m}^M {\sqrt {{\beta _m}{\beta _p}} E\left[ {\sqrt {{\upsilon ^{\left( {{B_p} + {B_m}} \right)}}} } \right]E\left[ {\sum\limits_{{n_1} = 1}^{N/M} {\left| {{f_{{\rm{ru}},p,{n_1}}}} \right|} \sum\limits_{{n_2} = 1}^{N/M} {\left| {{f_{{\rm{ru}},m,{n_2}}}} \right|} } \right]} } \\
+ \sum\limits_{m = 1}^M {{\beta _m}E\left[ {{\upsilon ^{{B_m}}}} \right]E\left[ {{{\left( {\sum\limits_{n = 1}^{N/M} {\left| {{f_{{\rm{ru}},m,n}}} \right|} } \right)}^2}}\right]} \end{array} \right)\\
&\overset{(b)}{=} \rho\left( \begin{array}{l}
{\beta _0}\omega \exp \left( { \!-\! E\left( {{B_0}} \right)\left( {1 \!-\! v} \right)} \right)\\ \!+\! \sqrt {{\beta _0}\omega } \frac{{\pi N}}{{2M}}\sum\limits_{m = 1}^M {\sqrt {{\beta _m}} \exp \left( { \!-\! \left( {E\left({{B_0}} \right)  \!+\! E\left({{B_m}} \right)} \right)\left( {1 \!-\! \sqrt v } \right)} \right)} \\
+ \frac{{\pi {N^2}}}{{4{M^2}}}\sum\limits_{m = 1}^M {\sum\limits_{p = 1,p \ne m}^M {\sqrt {{\beta _m}{\beta _p}} \exp \left( { - \left( {E\left( {{B_p}} \right) + E\left( {{B_m}} \right)} \right)\left( {1 - \sqrt v } \right)} \right)} } \\ + \sum\limits_{m = 1}^M {{\beta _m}\exp \left( { - E\left( {{B_m}} \right)\left( {1 - v} \right)} \right)\left( {\frac{N}{M} + \frac{N}{M}\left( {\frac{N}{M} - 1} \right)\frac{\pi }{4}} \right)}
\end{array} \right)\\
\end{aligned}
\end{array}
\end{equation}

Moreover, given that
\begin{equation}
E\left[ {\left( {\sum\limits_{n = 1}^{N/M} {| {{f_{{\rm{ru}},m,n}}} |} } \right)}^2\right]
\!=\!E\left[ {\sum\limits_{{n_1} = 1}^{N/M} {{{| {{f_{{\rm{ru}},m,{n_1}}}} |}^2}}  \!+\! \sum\limits_{{n_1} = 1}^{N/M}} \mathop{\sum\limits_{{n_2} = 1,}^{N/M}}_{{n_2} \ne {n_1}} {| {{f_{{\rm{ru}},m,{n_1}}}} |} | {{f_{{\rm{ru}},m,{n_2}}}} |  \right]
\!=\! {\frac{N}{M} \!+\! \frac{\pi N}{4M}\left( {\frac{N}{M} \!-\! 1} \right)},
\end{equation}
\begin{equation}
E\left( {\left| {{f_{{\rm{bu}}}}} \right|^2} \right)=1, 
E\left( {\left| {{f_{{\rm{bu}}}}} \right|} \right)=\frac{{\sqrt \pi  }}{2},
\end{equation}
\begin{equation}
\begin{aligned}
E\left( {{v^{{B_m}}}} \right) &= \sum\limits_{s = 0}^\infty  {{v^s}} \frac{{{{\left( {E\left( {{B_m}} \right)} \right)}^s}{e^{ - E\left( {{B_m}} \right)}}}}{{s!}} 
= {e^{ - E\left( {{B_m}} \right)(1 - v)}}, m=1, 2,.., M,
\end{aligned}
\end{equation}
\begin{equation}
\begin{aligned}
E\left( {{v^{{B_0}}}} \right)&={e^{ - E\left( {{B_0}} \right)(1 - v)}},\\
E\left( {{\sqrt{v^{{B_0}}}}} \right) &= {e^{ - E\left( {{B_0}} \right)(1 - \sqrt{v})}},\\
E\left( {{\sqrt{v^{{B_m}}}}} \right) &= {e^{ - E\left( {{B_m}} \right)(1 - \sqrt{v})}}, m=1,2,...,M,
\end{aligned}
\end{equation}
by substituting (41-44) into the expression after $\overset{(a)}{=}$ (\ref{Erzero}), we obtain the expression after $\overset{(b)}{=}$ in (\ref{Erzero}).
By substituting (\ref{blocker0}), (\ref{blockerm}), (\ref{PLm}) and (\ref{PL0}) into the expression after $\overset{(b)}{=}$ in (\ref{Erzero}), we present $E\left( {{\gamma _\varnothing}}\right)$ as a function of $M$, $h$, $\lambda_{\rm{B}}$, $R_{\rm{B}}$, and $v$, which is shown in (24).


\end{document}